\documentclass[journal, twocolumns]{IEEEtran}

\usepackage[utf8]{inputenc}
\usepackage{epsfig, amsthm, amsmath, amssymb, amsfonts, subfigure,color}

\usepackage{cite}
\usepackage{graphicx}
\usepackage{stfloats}
\usepackage[table,xcdraw]{xcolor}
\usepackage{tikz}
\usepackage{enumitem}
\usepackage{url}
\usepackage{array}
\usepackage{multirow}
\usepackage{booktabs} 
\usepackage{siunitx}  

\usepackage{xfrac}
\usepackage{nicefrac}
\usepackage{psfrag}
\usepackage{times}
\usepackage{balance} 
\usepackage{amssymb}
\usepackage{amsfonts}
\usepackage{bm}
\usepackage[bottom]{footmisc}
\usepackage{mwe}
\usepackage{color}
\usepackage{verbatim}
\usepackage{graphicx}
\usepackage{booktabs,multirow}
\usetikzlibrary{patterns,arrows}
\usepackage{epstopdf}

\usepackage{algorithm}
\usepackage{algorithmicx}
\usepackage{algpseudocode}
\usepackage{tabularx}

\usepackage[acronym]{glossaries}

\newacronym{iid}{i.i.d.}{independent and identically distributed}

\newacronym{lqr}{LQR}{linear quadratic regulator}

\newacronym{bs}{BS}{base station}

\newacronym{resnet}{ResNet}{residual network}

\newacronym{awgn}{AWGN}{additive white Gaussian noise}

\newacronym{gpr}{GPR}{Gaussian process regression}

\newacronym{inf}{InF-SH}{Indoor Factory}

\newacronym{std}{std}{standard deviation}

\newacronym{ofdma}{OFDMA}{orthogonal frequency division multiple access}

\newacronym{prb}{PRB}{physical resource block}

\newacronym{jepa}{JEPA}{joint-embedding predictive architecture}

\newacronym{i-jepa}{I-JEPA}{Image-Joint-Embedding Predictive Architecture}

\newacronym{v-jepa}{V-JEPA}{Video-Joint-Embedding Predictive Architecture}

\newacronym{ts-jepa}{TS-JEPA}{time-series joint-embedding predictive architecture}

\newacronym{mcs}{MCS}{modulation and coding schemes}

\newacronym{snr}{SNR}{signal-to-noise ratio}

\newacronym{cdf}{CDF}{cumulative distribution function}

\newacronym{pdf}{PDF}{probability distribution function}

\newacronym{ctde}{CTDE}{centralized training with decentralized execution}

\newacronym{mse}{MSE}{mean squared error}

\newacronym{los}{LoS}{line-of-sight}

\newacronym{nlos}{NLoS}{non-line-of-sight}

\newacronym{pca}{PCA}{principal component analysis} 

\newacronym{sgd}{SGD}{stochastic gradient descent} 

\newacronym{nrmse}{NRMSE}{normalized root mean squared error} 

\newacronym{cnns}{CNNs}{convolution neural networks} 

\newacronym{ebms}{EbMs}{energy-based models} 

\newacronym{jea}{JEA}{joint-embedding architecture} 

\newacronym{ga}{GA}{Generative Architecture} 

\newacronym{vit}{ViT}{Vision Transformer} 

\newacronym{ema}{EMA}{exponential moving average} 

\newacronym{mlp}{MLP}{Multi-layer Perceptron} 

\newacronym{cnn}{CNN}{Convolutional Neural Network } 

\newacronym{relu}{ReLu}{rectifier linear unit} 

\newacronym{mae}{MAE}{mean absolute error}

\newacronym{mape}{MAPE}{mean absolute percentage error}

\newacronym{nmae}{NMAE}{ normalized mean absolute error}

\newacronym{t-sne}{t-SNE}{t-Distributed Stochastic Neighbor Embedding}

\newacronym{urllc}{URLLC}{ultra-reliable low-latency communication}

\newacronym{mmtc}{mMTC}{massive machine-type communication}

\newacronym{ai}{AI}{artificial intelligence}

\newacronym{ml}{ML}{machine learning}

\newacronym{rnns}{RNNs}{recurrent neural networks}

\newacronym{lstm}{LSTM}{long short-term memory}

\newacronym{gru}{GRU}{gated recurrent unit}

\newacronym{aos}{AoS}{age of semantics}

\newacronym{aoi}{AoI}{age of information}

\newacronym{byol}{BYOL}{bootstrap your own latent}

\newacronym{sc}{SC}{Semantic Communication}

\newacronym{vae}{VAE}{variational auto-encoder}

\newacronym{simclr}{SimCLR}{simple framework for contrastive learning of visual representations}

\newacronym{bpp}{bpp}{bits per pixel}

\newacronym{arima}{ARIMA}{autoregressive integrated
moving average}

\newacronym{csi}{CSI}{channel state information}

\definecolor{color1}{RGB}{237, 191, 193}
\definecolor{color2}{RGB}{229, 153, 157}
\definecolor{color3}{RGB}{225, 123, 116}
\definecolor{color4}{RGB}{217,87,77}
\definecolor{color5}{RGB}{203,52,38}

\def\l{\left}
\def\r{\right}
\def\({\l(}
\def\){\r)}
\def\[{\l[}
\def\]{\r]}

\def\BibTeX{{\rm B\kern-.05em{\sc i\kern-.025em b}\kern-.08em
    T\kern-.1667em\lower.7ex\hbox{E}\kern-.125emX}}

\begin{document}

\title{Time-Series JEPA for Predictive Remote Control under Capacity-Limited Networks}

\author{
Abanoub M. Girgis,~\IEEEmembership{Student Member,~IEEE},
Alvaro~Valcarce,~\IEEEmembership{Senior Member,~IEEE},
and Mehdi~Bennis,~\IEEEmembership{Fellow,~IEEE}
\thanks{This work was supported in part by the European Union through the Project CENTRIC under Grant 101096379; in part by the RCF-Korea  (Semantics-Native Communication and Protocol Learning in 6G); and in part by the Research Council of Finland (former Academy of Finland) Project Vision-Guided Wireless Communication.}

\thanks{A. M. Girgis and M. Bennis are with the Center for Wireless Communications, University of Oulu, Oulu 90014, Finland (e-mail: abanoub.pipaoy@oulu.fi; mehdi.bennis@oulu.fi).}
\thanks{A. Valcarce is with Nokia Bell Labs, Massy, France (e-mail: alvaro.valcarce$\_$rial@nokia-bell-labs.com).}
}

\maketitle

\begin{abstract}
In remote control systems, transmitting large data volumes (e.g., images, video frames) from wireless sensors to remote controllers is challenging when uplink capacity is limited (e.g., RedCap devices or massive wireless sensor networks). 
Furthermore, controllers often need only information-rich representations of the original data. 
To address this, we propose a semantic-driven predictive control combined with a channel-aware scheduling to enhance control performance for multiple devices under limited network capacity. 
At its core, the proposed framework, coined Time-Series Joint Embedding Predictive Architecture (TS-JEPA), encodes high-dimensional sensory data into low-dimensional semantic embeddings at the sensor, reducing communication overhead. 
Furthermore, TS-JEPA enables predictive inference by predicting future embeddings from current ones and predicted commands, which are directly used by a semantic actor model to compute control commands within the embedding space, eliminating the need to reconstruct raw data. 
To further enhance reliability and communication efficiency, a channel-aware scheduling is integrated to dynamically prioritize device transmissions based on channel conditions and age of information (AoI).
Simulations on inverted cart-pole systems show that the proposed framework significantly outperforms conventional control baselines in communication efficiency, control cost, and predictive accuracy. 
It enables robust and scalable control under limited network capacity compared to traditional scheduling schemes.

%

\end{abstract}


\section{Introduction}
\label{Sec_Introd}
\IEEEPARstart{S}{emantic} communication has emerged as a critical enabler for the efficiency and scalability of next-generation 6G applications~\cite{9771334,10054510}, such as autonomous systems, smart cities, and immersive augmented reality~\cite{he20206g,ismail2022artificial,azuma1997survey}. 
In contrast to 5G technologies, where \gls{urllc} focuses on achieving high reliability and ultra-low latency at the cost of significant communication resources~\cite{8705373}, and \gls{mmtc} prioritizes massive connectivity with reduced quality-of-service~\cite{bockelmann2016massive}, semantic communication shifts the emphasis toward extracting and transmitting only the most relevant and meaningful features of the data. 
This paradigm aims to enhance both communication and control efficiency by reducing redundant transmissions and focusing on extracting and leveraging the intrinsic meaning embedded within the data.

A key aspect of semantic communication in control applications is the ability to represent high-dimensional sensor data (e.g., images or video frames) as low-dimensional semantic embeddings. 
These embeddings maintain the information necessary for downstream tasks while discarding redundancies and reducing the communication overhead.
As a result, efficient resource management for transmitting semantic embeddings becomes vital, particularly in limited wireless networks, to ensure timely and accurate control decisions for large-scale control systems.
This semantic-driven approach not only improves network scalability and reduces latency but also aligns closely with the growing need for intelligence at the edge, where devices transmit only distilled semantic information rather than raw data.
This shift in communication priorities motivates the integration of semantic representations with dynamic scheduling and machine learning techniques, as discussed in the following sections.

\subsection{Related Work}
Efficient scheduling is a fundamental requirement for scalable and reliable remote control in next-generation 6G wireless networks. 
Existing scheduling strategies, including round-robin~\cite{hespanha2007survey,schenato2007foundations}, opportunistic~\cite{han2017optimal,yu2019event,gatsis2015opportunistic}, stability-aware~\cite{girgis2021predictive}, and control-aware approaches~\cite{eisen2020control}, have been developed to allocate wireless resources while maintaining acceptable control performance. 
However, these approaches typically rely on high-dimensional data transmissions and are not designed to optimize the semantic efficiency of transmitted information. 
Moreover, their scalability is often limited by the increasing computational and communication overhead as the number of devices grows.

Recent advances in \gls{ml} offer promising alternatives by enabling predictive modeling of device dynamics and encoding raw sensory data into task-relevant representations. 
In particular, existing approaches in this domain can be broadly categorized into \gls{jea} and generative architectures.
The \gls{jea}s aim to learn representations such that embeddings of semantically similar inputs are mapped close together, while dissimilar ones are mapped apart. 
Popular examples include \gls{simclr}~\cite{chen2020simple} and \gls{byol}~\cite{grill2020bootstrap}.
Although the \gls{jea}s are computationally efficient and avoid pixel-level reconstruction, they are prone to representation collapse, where all inputs converge to a single embedding.
To mitigate this, strategies such as contrastive losses and momentum encoders have been proposed~\cite{guo2020bootstrap, chen2020improved}.

On the other hand, generative architectures, including \gls{rnns}-based models~\cite{connor1994recurrent,mandic2001recurrent}, \gls{lstm}~\cite{graves2012long}, \gls{gru}~\cite{dey2017gate}, \gls{cnns}~\cite{chung2014empirical}, and \gls{vae}~\cite{yu2019review}, reconstruct high-dimensional data from conditional inputs. 
These generative models are robust to representation collapse but often incur high computational cost and require large amounts of labeled data, making them less suitable for real-time remote control with limited wireless resources.

The \gls{jepa}~\cite{lecun2022path,assran2023self} offers a middle ground by predicting only relevant aspects of target data in a latent space, rather than reconstructing it in a raw high-dimensional space.
Specifically, the \gls{jepa}s seek to predict the embeddings of target high-dimensional data from the embeddings of input data and a conditional context, with a loss computed in the latent space, making it computationally efficient and scalable.
Applications of \gls{jepa}s have shown success in various domains, including image~\cite{assran2023self}, video~\cite{bardes2023v}, and audio~\cite{fei2023jepa}, demonstrating their potential for real-time control under limited network capacity.

\subsection{Contributions and Organization}
To address the critical challenge of remotely controlling multiple devices that share a limited wireless uplink to transmit high-dimensional sensory data, our key contributions are summarized as follows. 
\begin{itemize}
    \item We propose a novel \textit{\gls{ts-jepa}} that encodes high-dimensional sensory data at the device into low-dimensional semantic embeddings.
    The \gls{ts-jepa} captures the latent dynamics, enabling the prediction of future embeddings at the remote controller without reconstructing high-dimensional sensory data. 
    This significantly reduces communication overhead without compromising control performance. 

    \item On top of \gls{ts-jepa} embeddings, we train a \textit{semantic actor model} that directly maps low-dimensional semantic embeddings to control commands.
    This avoids the need to reconstruct high-dimensional data, efficiently reducing communication and computation costs.

    \item We develop a \textit{channel-aware scheduling} to dynamically prioritize devices for transmission based on channel conditions and the \gls{aoi}, ensuring that the most time-sensitive and reliable updates are transmitted under limited network capacity.

    \item Extensive simulations on inverted cart-pole systems demonstrate that the proposed framework achieves control performance comparable to conventional baselines while significantly reducing communication overhead. 
    Furthermore, the proposed approach scales effectively to multi-device scenarios, highlighting its practical applicability for remote control under limited network capacity.
\end{itemize}
The remainder of the paper is organized as follows. 
Section~\ref{Sec_System} describes the system model and the problem statement. 
In Section~\ref{Sec_solution}, we present the proposed semantic-driven predictive control, including \gls{ts-jepa} and the semantic actor model, combined with channel-aware scheduling.  
Section~\ref{Sec_results} presents simulation results that validate the effectiveness of the proposed approach in large-scale wireless control scenarios with limited network capacity.
Finally, Section~\ref{Sec_conclusion} concludes the paper and outlines future work.

\section{System Model}
\label{Sec_System}

\begin{figure}[t]
    \centering
    \includegraphics[width=0.5\textwidth]{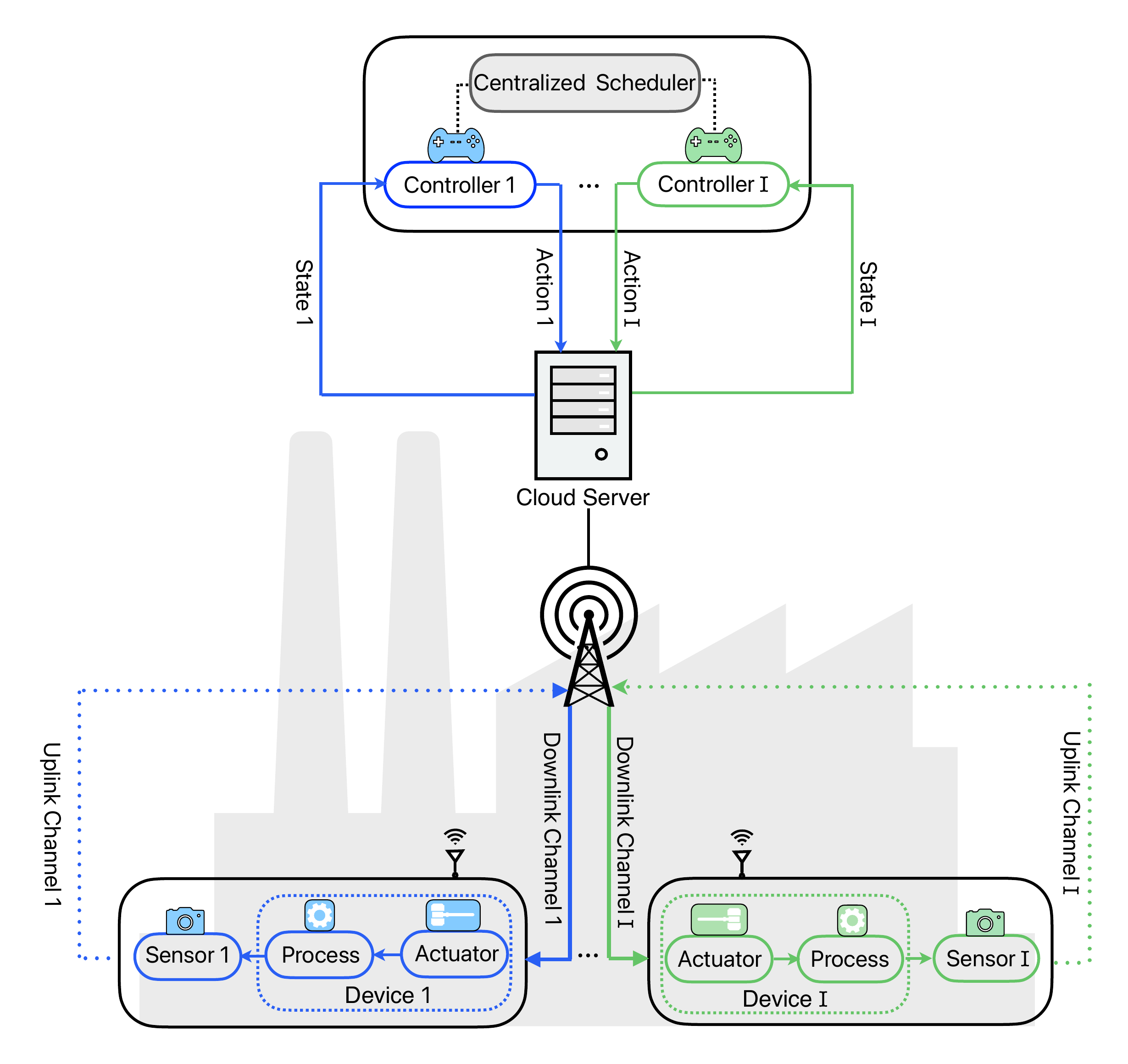} \\
    \caption{An illustration of wireless networked control systems in a smart factory.}
    \label{fig5_system_model}
\end{figure}

We consider a wireless networked control system, as shown in Fig.~\ref{fig5_system_model}, comprising multiple independent non-linear control systems that operate over shared wireless channels. 
Each control system includes a device, a sensor, and a remote controller.
The device integrates a non-linear dynamic \textit{process} that is being controlled along with an \textit{actuator} responsible for applying control commands to drive the process toward its desired state.
Each device is paired with a \textit{sensor} that periodically samples the device state to monitor its behavior.
The sampled state is transmitted via \textit{uplink transmission} to a high-computational remote controller.
Upon receiving the device state, the remote controller computes the appropriate control command and transmits it back to the actuator through downlink transmission to ensure that the process is steered toward its desired state. 
Further details on control and communication systems are discussed in the following sections.

\subsection{Control System}
\label{Sec5_A1}

We consider a wireless networked control system consisting of multiple independent devices, each equipped with a sensor and an actuator, communicating with a remote controller over wireless channels.
Each device is indexed by $ i\in \mathcal{I}$ with cardinality $|\mathcal{I}|$.
The sensor periodically samples the device's $p$-dimensional state vector at a fixed sampling rate $\tau_o$, where the sampled state for device $i \in \mathcal{I}$ at time $t = k \tau_o$ is represented as $\mathbf{x}_{i,k} \in \mathbb{R}^{p}$.   
Upon receiving the sampled state $\mathbf{x}_{i,k}$, the remote controller computes a target $q$-dimensional control command vector given as $ \mathbf{u}_{i,k} \in \mathbb{R}^{q}$.

The calculated control command is then transmitted over ideal channels to the actuator, which applies it to the corresponding device. 
The state evolution of device $i$ follows the discrete-time non-linear dynamics given as~\cite{ogata2010modern}
\begin{align}
    \label{eq5_non_linear_evol}
    \mathbf{x}_{i,k+1} = \mathbf{f}_{i} \left( \mathbf{x}_{i,k} , \mathbf{u}_{i,k} \right) + \mathbf{n}_{s,k},
\end{align}
where $\mathbf{n}_{s,k} \in \mathbb{R}^{p}$ is a process noise vector modeled as an \gls{iid} Gaussian random variables with zero mean and variance $N{_s}$.
The non-linear function $\mathbf{f}_{i} : \mathbb{R}^{p}  \times \mathbb{R}^{q} \rightarrow \mathbb{R}^{p}$ maps the current state and control command to the next state, representing the non-linear dynamics of the device $i$.

To ensure optimal control, the remote controller computes the target control command utilizing the non-linear control policy to solve the optimization problem formulated as~\cite{khalil2002nonlinear}
\begin{align}  
\label{eq5_Optimization_prob}
&\mathbf{u}^{*}_{i,k} = \underset{\mathbf{u}_{i,k}}{\arg \min} \; \;  \mathcal{J}(\mathbf{x}_{k}, \mathbf{u}_{k}) \\
& \qquad \; \; \text{subject to:} \; \eqref{eq5_non_linear_evol}, \nonumber \\
&\qquad \qquad \qquad \quad \mathbf{u}_{min} \leq \mathbf{u}_{i,k} \leq \mathbf{u}_{max}, \nonumber 
\end{align}
where the quadratic cost function is defined as 
\begin{align}
    \mathcal{J}(\mathbf{x}_{i,k}, \mathbf{u}_{i,k}) =  \frac{1}{2} \sum_{k = 0}^{K} \| \mathbf{x}_{i,k} - \mathbf{x}_{d} \|^{2}_{F} + \mathbf{u}^{\mathsf{T}}_{i,k} \, \mathbf{R} \, \mathbf{u}_{i,k},
\end{align}
where $\| \cdot \|_{F}$ is the Frobenius norm capturing the state deviation from the desired state $\mathbf{x}_{d}$, while $\mathbf{R}$ is a positive definite matrix penalizing control effort. 
Solving the optimization problem in~\eqref{eq5_Optimization_prob} through dynamic programming~\cite{fadali2012digital,bertsekas2012dynamic} yields the target control commands.
As depicted in Fig.~\ref{fig5_system_model}, each device's control loop is based on wireless channels for state transmission to ensure successful control performance. 
Since the device dynamics is inherently unstable, uplink transmission failures prevent the application of appropriate control commands, causing the state $\mathbf{x}_{i,k}$ to diverge to infinity as $k \rightarrow \infty$.

\subsection{Wireless Communication System}
We consider a sparse clutter and high base station \gls{inf} scenario, where regularly structured devices are randomly distributed within the coverage area of the base station.
As shown in Fig.~\ref{fig5_system_model}, remote controllers are assumed to be located on a distant cloud server with negligible communication delay between the cloud server and the base station.
The communication between the base station and the devices occurs through dedicated data and control channels. 
The data channels facilitate the exchange of device states and control commands, while the control channels manage state information such as scheduling requests and grants.

At each time step $t = k \tau_o$, the centralized scheduler at the base station dynamically manages the uplink channel access by issuing scheduling grants via error-free downlink control channels. 
These control channels operate without contention or collisions, ensuring reliable command delivery. 
The data channels follow a standard path loss and Rayleigh block fading model~\cite{goldsmith2005wireless}, where the channel gains for each device remain constant over the duration of $\tau_o$ but vary independently across different time intervals.
In the \gls{inf} scenario, the \gls{los} path loss between the $i$-th device and its remote controller is given as~\cite{etsi5138,9312675}
\begin{align}
    \label{eq_path_loss_Los}
    \mathrm{PL}^{\mathrm{LoS}}_{\mathrm{dB}} = 31.84 + 21.5 \log_{10} \left( D_{i}^{\mathrm{3D}} \right) + 19 \log_{10} \left( W_{c} \right), 
\end{align}
where $D_{i}^{\mathrm{3D}}$ denotes the 3-dimensional distance (in meters) between the $i$-th device and the remote controller, while $W_{c}$ represents the center frequency (in $\mathrm{GHz}$).

The probability of the channel being in a \gls{los} state at a given distance is modeled as 
\begin{align}
    \label{eq_Prob_Los}
    \mathbb{P}_{\mathrm{LoS}} = \text{exp} \left[ - \frac{D_{i}^{\mathrm{2D}}}{-\frac{D_{\mathrm{clutter}}}{\ln (1-\delta)} }. \frac{h_{\mathrm{BS}} - h_{i,R}}{ h_{c} - h_{i,R}} \right],
\end{align}
where $D_{i}^{\mathrm{2D}}$ is the 2-dimensional distance between the $i$-th device and its remote controller.
Additionally, $D_{\mathrm{clutter}}$, $\delta$, $h_{c}$, $h_{i,R}$, and $h_{\mathrm{BS}}$ denote the clutter size, clutter density, clutter height, antenna height of the $i$-th device, and the base station’s antenna height, respectively.
If the channel is classified as \gls{nlos}, the path loss between the $i$-th device and its remote controller is calculated as
\begin{align}
    \label{eq_path_loss_NLOS}
    \mathrm{PL}^{\mathrm{NLoS}}_{\mathrm{dB}} = \max \left( \mathrm{PL}_{\mathrm{dB}}, \mathrm{PL}^{\mathrm{LoS}}_{\mathrm{dB}} \right),
\end{align}
where the general path loss model $\mathrm{PL}_{\mathrm{dB}}$ in~\eqref{eq_path_loss_NLOS} is given by~\cite{etsi5138,9312675}
\begin{align}
    \label{eq_path_loss}
    \mathrm{PL}_{\mathrm{dB}}  = 33.63 + 21.9 \log_{10} \left( D_{i}^{3D} \right) + 20 \log_{10} \left( W_{c} \right), 
\end{align}
with a shadow fading \gls{std} of $4.0$. 

Here, we consider uplink data transmission from the $i$-th device to its corresponding remote controller at time $t = k \tau_{o}$ with fixed transmission power $P_{i}$, the received \gls{snr} at the base station is given as
\begin{align}
    \label{eq5_SNR}
    \gamma_{i,k} = 10^{-\frac{\mathrm{PL}^{\mathrm{NLoS}}_{\mathrm{dB}}}{10}} \frac{P_{i} \mid H_{i,k} \mid^{2} }{ N_{c}}, 
\end{align}
where $H_{i,k}$ denotes the Rayleigh flat-fading channel gain between the $i$-th device and the base station at time $k$, while $N_{c}$ represents the \gls{awgn} power.
The uplink channel capacity for device $i$ at time $k$ is given as
\begin{align}
    \label{eq_UL_Rate}
    R_{i,k} = W_{i} \log_{2} (1 + \gamma_{i,k}),
\end{align}
where $W_{i}$ is the allocated transmission bandwidth. 
Reliable communication is crucial to ensuring control performance; therefore, an outage transmission occurs when the uplink channel capacity falls below a predefined threshold $\bar{R}$, which guarantees the required transmission rate for stable control.
The outage probability, representing the likelihood of unsuccessful transmission, is given by
\begin{align}
    \label{eq_outage}
    \epsilon_{i,k} &= \mathbb{P} \left[ R_{i,k} < \bar{R} \right] \nonumber \\ &=  1 - \text{exp} \left[ -10^{\frac{\mathrm{PL}^{\mathrm{NLoS}}_{\mathrm{dB}}}{10}} \frac{ N_c }{P_{i}} \left(2^{ \frac{\bar{R}}{W_{i} } } -1\right)  \right], 
\end{align}
which follows from the \gls{cdf} of the exponential random variable.
This outage probability highlights the dependence on key factors such as transmission power $P_{i}$, allocated bandwidth $W_{i}$, and device-to-controller distance $D^{\mathrm{2D}}_{i}$.
Hence, efficient utilization of wireless resources is essential to minimize communication failures and ensure reliable uplink transmission, which is critical for maintaining robust control performance in wireless networked control systems.

\subsection{Problem Statement}
In the considered wireless networked control system, ensuring timely and reliable state updates from devices to remote controllers is vital for maintaining robust control performance. 
However, the presence of multiple devices that transmit high-dimensional states over limited wireless resources introduces a significant challenge.
This challenge introduces a critical trade-off between communication efficiency and control performance. 
Hence, effective scheduling and efficient resource allocation are essential to balance these competing objectives for maintaining control performance under limited network capacity.
Existing solutions primarily rely on time-triggered or round-robin scheduling, which struggles to adapt to dynamic network conditions and varying control requirements~\cite{hespanha2007survey}. 
While recent approaches have explored adaptive transmission based on control states and channel conditions~\cite{han2017optimal,eisen2020control,girgis2020predictive}, these approaches often lack scalability when managing multiple devices due to their dependence on the transmission of raw high-dimensional states.
Throughout this work, the terms high-dimensional state and frame are used interchangeably to reflect visual observations of the device state captured by the sensor.

To address these limitations, we introduce a novel \textit{semantic-driven predictive control} combined with a \textit{channel-aware scheduling} to enhance control performance for multiple devices operating under limited network capacity.
Unlike traditional approaches that rely on transmitting raw high-dimensional states, our proposed framework efficiently encodes these states into a low-dimensional embedding space, significantly reducing communication overhead without compromising control performance.
The proposed semantic-driven predictive control learns a mapping $\Psi(\cdot)$ that transforms the high-dimensional device state $\mathbf{x}_{i,k}$ at time $t = k \tau_{o}$ into a low-dimensional embedding $\mathbf{z}_{i,k} = \Psi(\mathbf{x}_{i,k})$.
This embedding space captures the essential latent dynamics, enabling the prediction of future embeddings $\mathbf{z}_{i,k+K_{p}}$ directly from the current embedding and a sequence of predicted control commands $( \tilde{\mathbf{u}}_{i,k}, \dots, \tilde{\mathbf{u}}_{i,k+K_{p}-1} )$.
Specifically, a second mapping $\mathcal{P}(\cdot)$ is introduced to predict future embeddings given as 
\begin{align}
    \mathbf{z}_{i,k+K_{p}} = \mathcal{P} \left( \mathbf{z}_{i,k}| \tilde{\mathbf{u}}_{i,k}, \dots, \tilde{\mathbf{u}}_{i,k+K_{p}-1} \right),
\end{align}
where $K_p$ represents the prediction horizon and $\tilde{\mathbf{u}}_{i,k}$ denotes the predicted control command of device $i$ at time $t = k \tau_{o}$. 
This predictive capability eliminates the need to reconstruct the raw high-dimensional state for control command computation. 
Instead, low-dimensional embeddings are directly leveraged to compute control commands, optimizing for downstream control tasks under limited communication resources.
By transmitting low-dimensional embeddings instead of high-dimensional states, the proposed semantic-driven predictive control significantly reduces wireless communication overhead, allowing more devices to efficiently share limited wireless resources without compromising control performance. 
To further enhance reliability and communication efficiency, a channel-aware scheduling is integrated with the semantic-driven predictive control.  
This scheduling approach dynamically selects devices for transmission based on their channel conditions and \gls{aoi}, ensuring that the most critical updates are prioritized for transmission.
By balancing channel quality and information freshness, the channel-aware scheduling maximizes resource utilization, minimizes transmission failures, and effectively mitigates outdated control information at remote controllers.

\section{Semantic-driven Predictive Control with Channel-aware Scheduling}
\label{Sec_solution}

To address the key challenges in the wireless networked control system, this section introduces a novel semantic-driven predictive control combined with channel-aware scheduling. 
The proposed framework leverages a \gls{ts-jepa} enhanced with a semantic actor model to improve communication efficiency, reduce latency, and maintain robust control performance in large-scale devices under limited wireless resources. 

\begin{figure}[t]
    \centering
   \includegraphics[width=0.49\textwidth]{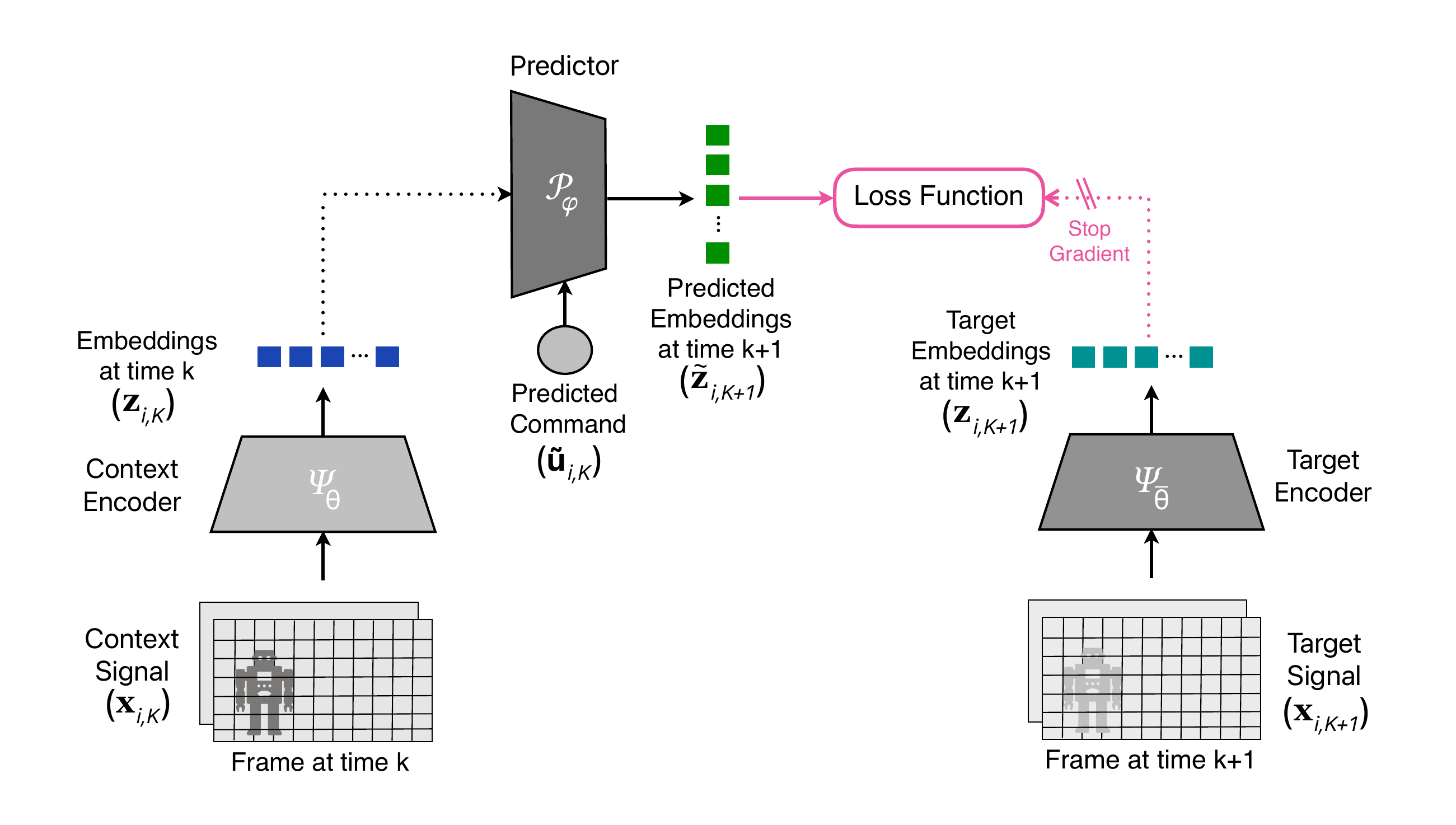} \\
   \caption{Time-series joint-embedding predictive architecture (TS-JEPA) to encode high-dimensional states into low-dimensional semantic embeddings and predict the future semantic embeddings.}
    \label{fig_TS_JEPA}
\end{figure}

\subsection{Time-Series Joint-Embedding Predictive Architecture}
\label{Ch5_1}
The proposed time-series joint-embedding predictive architecture (\gls{ts-jepa}) is a self-supervised learning technique designed to address the challenges in the wireless network control system by efficiently encoding high-dimensional device states into low-dimensional embeddings while enabling accurate prediction of future embeddings.
The key components of \gls{ts-jepa}, as illustrated in Fig.~\ref{fig_TS_JEPA}, are described below.
\begin{enumerate}
    \item \textit{Context:} At time step $t= k \tau_{o}$, the context encoder $\Psi_{\theta}(\cdot)$, parameterized by a set $\theta$ of learnable parameters, processes the high-dimensional state $\mathbf{x}_{i,k}$ and maps it to a low-dimensional embedding $\mathbf{z}_{i,k} = \Psi_{\theta}(\mathbf{x}_{i,k})$.  

    \item \textit{Targets:} For a defined prediction horizon $K_{p}$, the target encoder $\Psi_{\bar{\theta}} (\cdot)$ processes a sequence of future device states $( \mathbf{x}_{i,k+1}, \dots, \mathbf{x}_{i,k+K_{p}} )$ to generate their corresponding low-dimensional embeddings $( \mathbf{z}_{i,k+1}, \dots, \mathbf{z}_{i,k+K_{p}} )$. 

    \item \textit{Predictions:} The predictor $\mathcal{P}_{\varphi}$, parameterized by a set  $\varphi$ of learnable parameters,  captures the non-linear embedding evolution to predict target embedding directly from the current embedding and predicted control commands.
    Specifically, the predictor infers the future embeddings based on current embeddings and predicted control commands $( \tilde{\mathbf{u}}_{i,k}, \dots, \tilde{\mathbf{u}}_{i,k+K_{p}-1} )$, solving an auto-regressive task as 
    \begin{align}
        \left( \tilde{\mathbf{z}}_{i,k+1}, \dots, \tilde{\mathbf{z}}_{i,k+K_{p}} \right) = \mathcal{P}_{\varphi} \left( \mathbf{z}_{i,k}| \tilde{\mathbf{u}}_{i,k}, \dots, \tilde{\mathbf{u}}_{i,k+K_{p}-1} \right), \label{eq_pred5}
    \end{align}
    where $\tilde{\mathbf{z}}_{i,k}$ is the predicted embedding at time $k$. 

    \item \textit{Loss:} To train \gls{ts-jepa}, a cosine similarity loss function is employed to align the predicted embeddings with their target embeddings. 
    The context encoder's and predictor's parameters $\theta$ and $\varphi$ are jointly learned through gradient-based optimization as
    \begin{align}
         \label{eq_opt_problem}
         \underset{ \theta, \varphi }{\arg \min}  \; \; \frac{1}{K_{s}} \sum_{k = 1}^{K_s} \frac{ \langle  \tilde{\mathbf{z}}_{i,k+1}, \mathbf{z}_{i,k+1} \rangle}{ \|  \tilde{\mathbf{z}}_{i,k+1} \|_{2}.  \|  \mathbf{z}_{i,k+1}) \|_{2}},     
    \end{align}
    while the target encoder's parameters $\bar{\theta}$ are updated, at each training step, via an \gls{ema} of the context encoder's parameters to ensure stable training.
    The target encoder's parameters update rule is given as 
    \begin{align}
        \label{eq_weights_update}
         \bar{\theta} \leftarrow \eta \bar{\theta}  + (1 - \eta) \theta. 
    \end{align} 
    where $\eta \in [0,1 ]$ is a decay rate that regulates the weight update rate~\cite{grill2020bootstrap}. 
\end{enumerate}
Specifically, the target encoder mirrors the context encoder’s architecture and shares identical parameters at initialization.
Since the target encoder provides labels for training the context network, its gradients are blocked through its branch to prevent representation collapse, and its weights are updated using an \gls{ema} of the context encoder's parameters in~\eqref{eq_weights_update}.
This iterative process gradually enhances the context encoder's ability to produce meaningful low-dimensional embeddings from which the predictor infers future embeddings. 
Once trained, the proposed \gls{ts-jepa} maps high-dimensional device states to low-dimensional embeddings, enabling auto-regressive prediction of future embeddings by conditioning on predicted control commands. 
%
%

\begin{algorithm}[ht]
\caption{Time-series JEPA Algorithm}
\label{alg:TSJEPA}
\begin{algorithmic}[1]
\Require High-dimensional stat sequence $\{\mathbf{x}_{i,k}\}_{k=1}^{K_s}$, predicted control commands $\{\tilde{\mathbf{u}}_{i,k}\}_{k=1}^{K_s}$
\Ensure Trained parameters $\theta$ and $\varphi$

\State \textbf{Initialize:} Context encoder $\Psi_{\theta}$, predictor $\mathcal{P}_{\varphi}$ with random weights
\State Set target encoder $\Psi_{\bar{\theta}} \gets \Psi_{\theta}$
\State Set EMA decay rate $\eta \in [0,1]$

\For{each training step $k = 1$ to $K_s$}
    \State \textbf{Context Encoding:} $\mathbf{z}_{i,k} \gets \Psi_{\theta}(\mathbf{x}_{i,k})$
    
    \State \textbf{Target Encoding:}
    \For{$j = 1$ to $K_p$}
        \State $\mathbf{z}_{i,k+j} \gets \Psi_{\bar{\theta}}(\mathbf{x}_{i,k+j})$
    \EndFor

    \State \textbf{Prediction:}
    \State Predict future embeddings as in eq.~\eqref{eq_pred5}.

    \State \textbf{Loss cosine similarity computation}
    
    \State \textbf{Gradient Update:} Update $\theta$ and $\varphi$ using gradient descent as in eq.~\eqref{eq_opt_problem}
    
    \State \textbf{EMA Update of Target Encoder} as in eq.~\eqref{eq_weights_update}  
\EndFor
\State \textbf{Return}  Trained $\Psi_{\theta}$ and $\mathcal{P}_{\varphi}$
\end{algorithmic}
\end{algorithm}

In a nutshell, the proposed \gls{ts-jepa} for remote monitoring in wireless networked control systems operates through two distinct phases. 
\begin{enumerate}
    \item \textit{Training Phase:} During this phase, the devices transmit their high-dimensional states to remote controllers for control command computation. 
    Meanwhile, the proposed \gls{ts-jepa} is trained at the base station using a collected dataset, denoted as $\mathcal{D}_{s} = \{ \mathbf{x}_{i,k}, \mathbf{u}_{i,k} \}_{k=1}^{K_{s}}$, where $K_{s}$ represents the number of consecutive states and their corresponding control commands.
    The training process, detailed in Algorithm~\ref{alg:TSJEPA}, iteratively optimizes the context encoder and predictor models until convergence, ensuring that the learned low-dimensional embeddings effectively minimize prediction errors.  

    \item \textit{Inference Phase:} Once training is complete, the learned context encoder is deployed on the devices to encode high-dimensional states into low-dimensional embeddings. 
    Meanwhile, the learned predictor is deployed at the cloud server to predict future embeddings by conditioning current embeddings on predicted control commands. 
    This predictive capability enables the remote controller to predict latent dynamics, reducing reliance on frequent state transmissions and improving communication efficiency.
\end{enumerate}
While these low-dimensional embeddings, whether received from the context encoder or predicted using the predictor, efficiently reduce wireless overhead, directly computing control commands from these embeddings poses a challenge.
To address this challenge, a \textit{semantic actor model} is introduced to map the low-dimensional embeddings to calculated control commands.
This model effectively bridges the gap between the learned low-dimensional embeddings and control command computation.

\begin{figure}[t]
    \centering
   \includegraphics[width=0.49\textwidth]{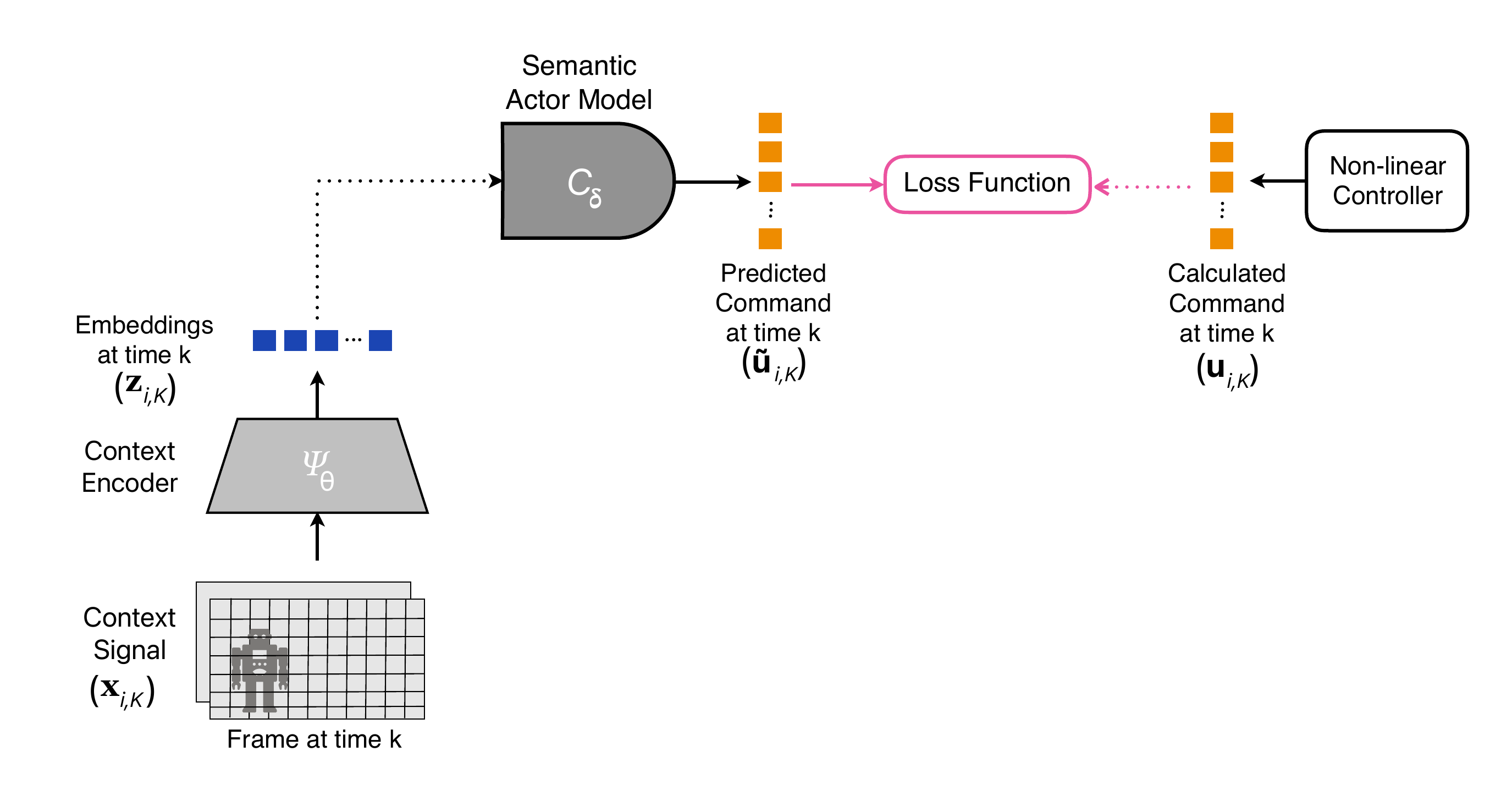} \\
   \caption{Semantic actor model to predict control commands from low-dimensional semantic embeddings.}
    \label{fig_Actor}
\end{figure} 

\subsection{Semantic actor model}
\label{Ch5_2}

Once the \gls{ts-jepa} is well trained, the context encoder encodes high-dimensional states into low-dimensional embeddings.
These embeddings are used to form a dataset $\mathcal{D}_{a} = \{ \mathbf{z}_{i,k}, \mathbf{u}_{i,k} \}_{k=1}^{K_a}$, , which contains $K_{a}$ consecutive low-dimensional embeddings and their corresponding control command.
To predict control commands directly from these low-dimensional embeddings, the \textit{semantic actor model} $\mathcal{C}_{\varepsilon}$, illustrated in Fig.~\ref{fig_Actor}, is trained using the dataset $\mathcal{D}_{a}$.
The semantic actor model minimizes the average squared $l_{2}$-distance between the predicted and target control commands, with its parameters optimized via gradient-based optimization as 
\begin{align}
     \label{eq_opt_problem_2}
      \underset{ \varepsilon }{\arg \min}  \; \; \frac{1}{K_{c}} \sum_{k = 1}^{K_c} \Vert \mathbf{u}_{i,k} - \tilde{\mathbf{u}}_{i,k} \Vert_{2}^{2},
\end{align}
where $\tilde{\mathbf{u}}_{i,k}$ is the predicted control command of device $i$ at time $k$ and $K_{c}$ denotes the number of consecutive embeddings and their corresponding control commands used for training.

In a nutshell, the remote control procedure employing the semantic actor model consists of two main phases.   
\begin{enumerate}
    \item \textit{Training Phase:} During this phase, the devices transmit their high-dimensional states to remote controllers for control command computation. 
    Meanwhile, the proposed semantic actor model $\mathcal{C}_{\varepsilon}$ is trained at the base station using the dataset $\mathcal{D}_{a}$ until convergence. 

    \item \textit{Inference Phase:} Once the semantic actor model is well trained, it predicts control commands directly from low-dimensional embeddings.
    These embeddings may be received from the context encoder or predicted using the predictor in the proposed \gls{ts-jepa}. 
    This predictive capability ensures continuous control under limited wireless resources or adverse channel conditions.
\end{enumerate}
The proposed semantic-driven predictive control integrates \gls{ts-jepa} with a semantic actor model to ensure robust control performance while minimizing communication overhead in large-scale wireless control deployments. 
Although the proposed framework efficiently leverages low-dimensional embeddings to improve communication efficiency, scalability, and control accuracy under limited wireless resources, optimizing wireless resource utilization remains crucial in large-scale deployments.
To further improve communication efficiency, a channel-aware scheduling is introduced.
This scheduling approach dynamically selects devices for transmission based on their channel conditions and \gls{aoi}, ensuring that the most critical updates are prioritized for uplink transmission.

\subsection{Channel-aware scheduling}
\label{Ch5_3}

The proposed channel-aware scheduling is designed to dynamically prioritize devices for uplink transmission by evaluating both their channel conditions and \gls{aoi}. 
This scheduling approach ensures that the most critical updates are prioritized for transmission, maximizing communication efficiency and improving control performance in large-scale deployments.
The primary objective is to design a centralized scheduler that enhances communication efficiency while minimizing outdated control information. 
To achieve this, we formulate an optimization problem that jointly maximizes a weighted sum of the probability of successful transmission while minimizing \gls{aoi} across all devices while satisfying the resource allocation and transmission reliability constraints given as 
\begin{subequations}
\label{eq5_opt} 
 \begin{gather} 
 \label{eq5_opt_1} 
 \underset{\hspace{0pt} \[\alpha_{1,k} \cdots \alpha_{I,k}\]}{\text{Maximize}} \; \sum^{I}_{i=1} \omega_{1} \, \mathbb{P} \left[ \xi_{i,k} =1\vert H_{i,k}, P_{i}, \alpha_{i,k} \right] - \omega_{2} \, \log \left( \beta_{i,k}\right) \\
 \hspace{-90pt} \text{subject to:}\quad \alpha_{i,k} \in \{ 0, 1 \}, \qquad \forall i,k \label{eq5_opt_2} \\
 \hspace{-35pt} \sum_{i=1}^{I} \alpha_{i,k} \leq J \qquad \quad \forall k \label{eq5_opt_3} \\
 \hspace{-35pt} \gamma_{i,k} \geq \alpha_{i,k} \gamma_{th} \qquad  \forall i,k \label{eq5_opt_4} 
 \end{gather}
\end{subequations} 
where $\omega_{1}$ and $\omega_{2}$ are positive weighting hyperparameters that balance the trade-off between ensuring reliable transmission and prioritizing devices with outdated information. 
The constraints in~\eqref{eq5_opt_2}-\eqref{eq5_opt_3} ensure feasible scheduling variables, allowing at most $J$ devices to be scheduled at each time. 
Note that the number of available resource blocks 
$J$ is strictly less than the total number of devices $I$, reflecting the resource-constrained nature of the system.
The constraint in~\eqref{eq5_opt_4} guarantees reliable transmission by ensuring that the received \gls{snr} for any scheduled device satisfies a threshold $\gamma_{th}$.

The probability of successful transmission is defined as 
\begin{align}
    \mathbb{P} \left[ \xi_{i,k} =1 \vert H_{i,k}, P_{i}, \alpha_{i,k} \right] = \alpha_{i,k} (1 - \epsilon_{i,k}),
\end{align}
where $\xi_{i,k} \in \{ 0, 1 \}$ represents the transmission event of the device $i$ at time $k$ and $\alpha_{i,k} \in \{ 0,1 \}$ is the scheduling decision of the device $i$ at time $k$.
Meanwhile, the \gls{aoi} of the device $i$ at time $k$, representing the time elapsed since the device's most recent update, quantifies the freshness of the received information given as~\cite{kosta2017age}
\begin{align}
    \label{eq5_AoI}
    \beta_{i,k} = \left\{ \begin{array}{ll}  1 + \beta_{i,k-1}, &\text{if} \; \alpha_{i,k} = 0, \\  1, & \text{otherwise}.
 \end{array}
 \right.
\end{align}

To obtain optimal scheduling decisions, each device is assigned a cost that combines its transmission success probability and \gls{aoi}.
This cost is defined as 
\begin{align}
    \label{eq5_scoreE}
    S_{i} = \omega_{1} \left( 1 - \epsilon_{i,k} \right) - \omega_{2} \log \left( \beta_{i,k} \right).
\end{align}
The optimal scheduling decisions are then determined through the channel-aware scheduling algorithm~\ref{alg:scheduling}.
By jointly considering transmission reliability and \gls{aoi}, the proposed channel-aware scheduling effectively balances communication efficiency and control performance. 
This approach dynamically allocates resources to devices that require urgent updates, while ensuring reliable transmission.
Hence, the proposed scheduling mitigates the risk of outdated control information and maximizes network resource utilization in large-scale deployments.

\begin{algorithm}  
\caption{Channel-Aware Scheduling Algorithm}
\label{alg:scheduling}
\begin{algorithmic}[1]
\State \textbf{Initialization:} 
\State Set weights $\omega_{1}$ and $\omega_{2}$.
\State Initialize scheduling variables $\alpha_{i,k} = 0, \quad \forall i \in \{1, 2, \ldots, I\}$.
\State Initialize candidate set $\mathcal{S} = \emptyset$.

\For{\textbf{each} device $i = 1$ to $I$}
    \State Observe the previous AoI $\beta_{i,k-1}$ and compute the updated AoI: $ \beta_{i,k} = \alpha_{i,k} + (1 - \alpha_{i,k})(1 + \beta_{i,k-1})$
    \State Compute cost: $[
    S_{i} = \omega_1 (1 - \epsilon_{i,k}) - \omega_2 \log(\beta_{i,k}) ]$
    
    \If{ $\gamma_{i,k} \geq \gamma_{th}$ }
        \State Add device $i$ to candidate set $\mathcal{S}$
    \EndIf
\EndFor

\State Sort devices in $\mathcal{S}$ in \textbf{descending order} of $S_i$

\State Select the top $J$ devices from the sorted list.

\For{each selected device $i \in \mathcal{S}$ (top $J$)}
    \State Set $\alpha_{i,k} = 1$
\EndFor

\State \textbf{Return} $\alpha_{i,k}$ for all devices.

\end{algorithmic}
\end{algorithm}

\section{Simulation Results}
\label{Sec_results}

To evaluate the performance of the proposed semantic-driven predictive control with channel-aware scheduling, we conducted extensive simulations on inverted cart-pole systems in a large-scale deployment.
The inverted cart-pole system is selected because of its inherent instability and its strong dependence on timely and efficient communication between sensors and controllers to maintain robust control performance.
%
%
In our simulation setup, each inverted cart-pole system's state is represented by an RGB frame captured at a fixed sampling interval of $ \tau_{o} =  1\, \mathrm{ms}$  over a time interval of $100$ time steps.
This sampling rate is selected to ensure capturing meaningful temporal dynamics for effective predictive embeddings learning, as detailed later.
The corresponding control command applies a horizontal force to the cart to maintain control performance.
The control commands are generated based on a non-linear control policy with predefined control limits of $u_{max} = 20 \, \mathrm{N}$, and $u_{min} = -20 \, \mathrm{N}$~\cite{florian2007correct,morasso2019stabilization}.

\subsection{Data Generation and Training}

To train the proposed semantic-driven predictive control, we generated two distinct datasets: one for the \gls{ts-jepa} and another for the semantic actor model. 
The \gls{ts-jepa} dataset $\mathcal{D}_{s}$ includes $200$ training and $40$ testing trajectories, each consisting of RGB frames paired with their corresponding control commands.
Meanwhile, the semantic actor model’s dataset $\mathcal{D}_{a}$ comprises $100$ training and $20$ testing trajectories, each containing pairs of low-dimensional embeddings and their corresponding control commands.

The weight parameters of the proposed \gls{ts-jepa} are trained by minimizing the cosine similarity loss function defined in~\eqref{eq_opt_problem}. 
The proposed \gls{ts-jepa} is trained using the hyperparameters listed in Table~\ref{Table_2},  with a learning rate that decays by a factor of $0.99$ every $20$ epochs. 
Similarly, the semantic actor model is trained by minimizing the \gls{mse} loss function defined in~\eqref{eq_opt_problem_2}, using the hyperparameters in Table~\ref{Table_3}. 
To mitigate overfitting, we apply early stopping based on validation performance. 
Each experiment is repeated five times to enhance statistical reliability, and the best results are reported.

The semantic-driven predictive control was implemented and trained on a NVIDIA Tesla V100-PCIE-16GB GPU-accelerated platform to handle computational complexity efficiently.
The \gls{ts-jepa} encoder adopts a deep convolutional \gls{resnet} architecture with layers of $64$, $128$, and $256$ neurons, each followed by batch normalization and \gls{relu} activation.
The \gls{ts-jepa} predictor is structured as an \gls{mlp} with a hidden layer of $1024$ neurons and an output layer of $256$ neurons. 
The semantic actor model is implemented as an \gls{mlp} with two hidden layers containing $1024$ and $256$ neurons, respectively, each followed by a \gls{relu} activation function.

To improve the robustness of the proposed \gls{ts-jepa} by enhancing spatial diversity, the RGB frames follow a series of pre-processing steps of image augmentation inspired by self-supervised learning approaches~\cite{grill2020bootstrap,chen2020simple}:
\begin{itemize}
    \item \textit{Color Jittering:} Randomly adjust brightness $(0.05)$, contrast $(0.1)$, saturation $(0.1)$, and hue $(0.05)$ across all pixels, applied in random order for each patch to enhance visual diversity.
    
    \item \textit{Color Dropping:} Converts frames to grayscale with a $0.05$ probability, replacing RGB frame intensity with the luma component.
    
    \item \textit{Normalization:} Each color channel is normalized by subtracting the mean values $\left[ 0.485, 0.456, 0.406 \right]$  and dividing by the standard deviations $\left[ 0.229, 0.224, 0.225 \right]$ to improve convergence.  
    
    \item \textit{Resizing:} Frames are resized to $64 \times 128$ using a $5 \times 5$ Gaussian kernel with a standard deviation randomly sampled from $\left[ 0.1, 0.2 \right]$ to reduce computational overhead while keeping key visual features.        
\end{itemize}
Additionally, control commands are pre-processed using z-score normalization to ensure stable \gls{ts-jepa} training.
During \gls{ts-jepa} testing, resizing and normalization are applied to the RGB frames to ensure consistency with the training phase.

To evaluate the performance of the proposed framework under wireless conditions, simulations are conducted under varying target \gls{snr} values: $\gamma_{\text{th}} \in \{ 5, 10, 20 \}\, \mathrm{dB}$. 
The remaining wireless parameters used in the simulation are detailed in Table~\ref{Table_1}.

\subsection{Evaluation Metrics}
To evaluate the performance of the proposed semantic-driven predictive control with channel-aware scheduling, five key metrics are employed: encoder performance, temporal consistency, prediction accuracy, control performance, and communication efficiency. 
\begin{enumerate}
    \item \textit{\textbf{Encoder Performance}}: The quality of the trained context encoder of the proposed \gls{ts-jepa} is evaluated using \gls{t-sne}. 
    As a powerful dimensionality reduction technique, \gls{t-sne} maps high-dimensional embeddings into a two-dimensional space while preserving their local structure. 
    This technique effectively reflects the encoder's ability to generate meaningful embeddings, where similar embeddings in the original high-dimensional space remain closely clustered in the reduced space.
    Improved clustering indicates stronger semantic embeddings and improved encoder performance.

    \item \textit{\textbf{Temporal and spatial consistency}}: is evaluated to quantify the relative change in state values across successive time steps. 
    This metric is computed using the \gls{mape} between consecutive frames, defined as
    \begin{align}
        \label{eq5_MAPE}
        \mathcal{N}^{P}_{i,k} = \frac{1}{p} \sum_{\upsilon=1}^{p} \left| \frac{\mathbf{x}_{i,k}(\upsilon) - \mathbf{x}_{i,k-1} (\upsilon)}{\mathbf{x}_{i,k-1}( \upsilon)} \right| \times 100 \%,
    \end{align}    
    where $\mathcal{N}^{P}_{i,k}$ represents the \gls{mape} of device $i$ at time $k$. 

    \item \textit{\textbf{Prediction Accuracy}}: The accuracy of the proposed semantic actor model in predicting control commands is evaluated using the \gls{nmae} between the predicted and ground truth control commands, defined as
    \begin{align}
        \label{eq5_NMAE}
        \mathcal{N}^{u}_{i,K_{p}} = \frac{\frac{1}{K_{P}} \sum_{k=K_{s}+1}^{K_{s}+K_{P}}\left| \tilde{u}_{i,k} - u_{i,k} \right| }{ \left| \max (u) - \min(u) \right|},
    \end{align}
    where $\mathcal{N}^{u}_{i,K_{p}}$ quantifies the normalized prediction error of the $i$-th device over the $K_{P}$ prediction steps during testing. 
    The normalization term ensures that the metric is independent of the control command range, providing a fair evaluation across varying system scales.

    \item  \textit{\textbf{Control Performance}}:
    The control performance is evaluated using a scoring function that rewards successful control outcomes based on the system’s ability to drive the cart’s position and pendulum angle to the desired position and angle. 
    The scoring function is defined as
    \begin{align}
        \label{eq5_scoring}
        \mathcal{R}_{i,k} = \left\{ \begin{array}{cc}
            1 & \left| x_{i,k} - x_{d} \right| \leq0.05 \, \& \, \left| \vartheta_{i,k} \right| \leq 0.05  \\
            0 & \text{otherwise},
        \end{array} \right.
    \end{align}
    where $\mathcal{R}_{i,k}$ represents the score value of the $i$-th device at time $k$, $x_{i,k}$ is the cart's location, $x_{d}$ is the desired cart position, and $\vartheta_{i,k}$ is the pendulum's angle. 

    \item \textit{\textbf{Communication Efficiency}}: is measured by the number of communication bits required for each device to transmit its state or corresponding embedding at each step.
\end{enumerate}

\subsection{Baseline Models}
To evaluate the performance of the proposed semantic-driven predictive control with channel-aware scheduling, we compare it with three control baselines and two scheduling baseline policies. 
\subsubsection{Control Baselines}: The proposed semantic-driven predictive control is evaluated against the following control baselines. 
\begin{enumerate}
    \item \textit{\textbf{Baseline 1: Optimal Control Policy}}
    \begin{itemize}
        \item In this policy, the remote controller receives the high-dimensional state from the device and computes the corresponding control commands using a non-linear control policy. 
        This policy optimally balances state deviation and control effort, ensuring robust control performance. 
        The calculated control command is then transmitted to the actuator for application. 
        This policy serves as a benchmark for optimal control performance under ideal conditions.
    \end{itemize}

    \item \textit{\textbf{Baseline 2: Supervised Learning Model}}
    \begin{itemize}
        \item This approach utilizes a supervised learning model to predict control commands directly from the high-dimensional state.
        The remote controller processes the received device state through the trained supervised model, which maps the device states to control commands.
        The predicted control command is then applied by the actuator. 
    \end{itemize}

    \item \textit{\textbf{Baseline 3: Generative Auto-encoder Model}} 
    \begin{itemize}
        \item The generative auto-encoder model employs an encoder-decoder structure.
        The encoder, deployed on the device, encodes the high-dimensional state into a low-dimensional representation before transmitting it to the remote controller. 
        The remote controller then decodes the received representations to reconstruct the device state, which is subsequently used to compute the control commands.

    \end{itemize}
\end{enumerate}

\subsubsection{Scheduling Baselines}

The proposed channel-aware scheduling is evaluated against the following scheduling baseline approaches.
\begin{enumerate}
    \item \textit{\textbf{Baseline 4: Round-Robin Scheduling}}
    \begin{itemize}
        \item In this approach, each device periodically transmits its high-dimensional state to the remote controller following a predefined repeating order.
        The remote controller computes the control command, which is transmitted to the actuator for application.
        When a device is unscheduled, its actuator applies the most recently received control command~\cite{hespanha2007survey,schenato2007foundations}.
    \end{itemize}

    \item \textit{\textbf{Baseline 5: Opportunistic Scheduling }}
    \begin{itemize}
        \item This scheduling approach exploits channel conditions to determine when a device should transmit its state. 
        When a device’s channel conditions are poor, it remains unscheduled, and its actuator applies the previously received control command.
        By leveraging channel variations, this approach aims to improve transmission reliability and reduce communication overhead.
        However, frequent state updates may be missed if channel conditions remain unfavorable~\cite{xu2013stability,liu2003framework}.
    \end{itemize}
    
\end{enumerate}
The combination of these baselines provides a comprehensive evaluation of the proposed framework. 
The control baselines test the effectiveness of the semantic-driven predictive control, while the scheduling baselines evaluate the ability of the channel-aware scheduler to prioritize critical updates under limited network capacity. 

\begin{table}
    \centering
    \resizebox{.65\columnwidth}{!}{\begin{minipage}[t]{.55\columnwidth}
    \caption{TS-JEPA hyperparameters}
    \label{Table_2}
\begin{tabularx}{1\linewidth}{p{2.6cm} p{1.2cm}}
    \toprule[1pt]
    \textbf{Hyperparameters}   &  \textbf{Values} \\  
    \cmidrule(r){1-1}   \cmidrule(l){2-2} 
      Learning rate    &  $2 \times 10^{-1}$  \\ 
      Batch size  & $256$  \\
      Epochs number    & $150$  \\ 
      Optimizer & $\mathrm{SGD}$  \\
      Weight decay  & $0.0004$  \\
      \gls{ema} decay rate & $0.99$ \\
    \bottomrule[1pt]
\end{tabularx}
\end{minipage}}
\end{table}

\begin{table}
    \centering
    \resizebox{.65\columnwidth}{!}{\begin{minipage}[t]{.55\columnwidth}
    \caption{Semantic actor hyperparameters}
    \label{Table_3}
\begin{tabularx}{1\linewidth}{p{2.5cm} p{1.35cm}}
    \toprule[1pt]
    \textbf{Hyperparameters}   &  \textbf{Values} \\  
    \cmidrule(r){1-1}   \cmidrule(l){2-2} 
      Learning rate    &  $6 \times 10^{-3}$  \\ 
      Batch size  & $200$  \\
      Epochs number    & $300$  \\ 
      Optimizer & $\mathrm{AdamW}$  \\
      Dropout   & $0.2$  \\ 
    \bottomrule[1pt]
\end{tabularx}
\end{minipage}}
\end{table}

\begin{table}[t]
    \centering
    \resizebox{.85\columnwidth}{!}{\begin{minipage}[t]{.7\columnwidth}
    \caption{System parameters}
    \label{Table_1}
\begin{tabularx}{1\linewidth}{p{2.1cm} p{1.1cm} p{1.75cm}} 
    \toprule[1pt]
    \textbf{Parameters}  & \textbf{Symbol} &  \textbf{Values} \\  
    \cmidrule(r){1-1}   \cmidrule(l){2-2} \cmidrule(l){3-3}
      Hall size  &  & $300 \times 150 \; \mathrm{m}^{2}$  \\ 
      Room height & & $6 \; \mathrm{m}$  \\
      BS height   & $h_{\mathrm{BS}}$  & $10.0 \; \mathrm{m}$  \\ 
      Device height  & $h_{i,R}$ & $1.5 \; \mathrm{m}$  \\
      Carrier frequency & $w_c$  & $3.75 \; \mathrm{GHz}$  \\ 
      Total bandwidth &   & $20 \; \mathrm{MHz}$  \\ 
      Clutter height & $h_{c}$  & $ 3 \mathrm{m} $  \\ 
      Clutter size & $D_{\mathrm{clutter}}$  & $ 2.0 \; \mathrm{m}$  \\ 
      Clutter density & $\delta $   & $ 60 \%$  \\  
      2D distance & $D_{i}^{\mathrm{2D}} $   & $ 50 \; \mathrm{m} $  \\ 
      Noise power & $N_{c} $   & $ -95 \; \mathrm{dB} $  \\  
    \bottomrule[1pt]
\end{tabularx}
\end{minipage}}
\end{table}

\begin{figure}
    \centering
    \subfigure[without image augmentation.\label{fig_Wo_Aug}]{\includegraphics[width=0.40\textwidth]{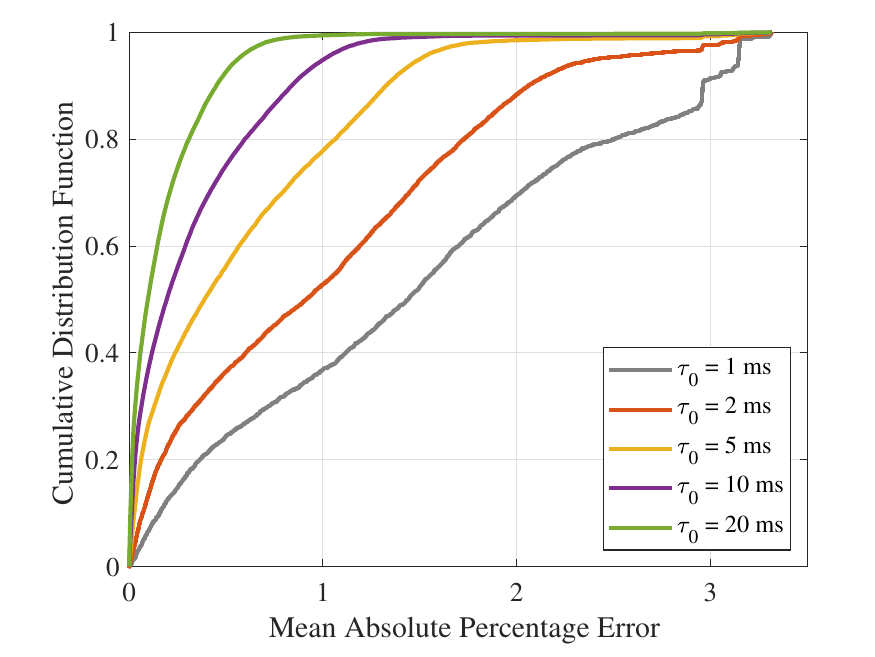}}
    \subfigure[with image augmentation.\label{fig_W_Aug}]{\includegraphics[width=0.40\textwidth]{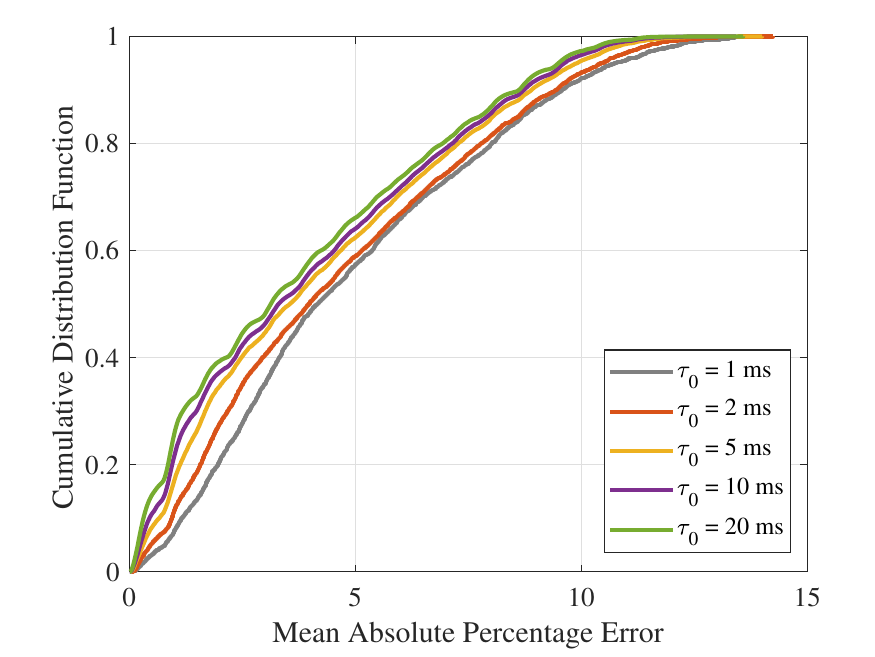}} 
   \caption{Cumulative distribution function of the mean absolute percentage error between consecutive frames in the training dataset under varying sampling rates: (a) without image augmentation and (b) with image augmentation.}
    \label{fig_data_Aug}
\end{figure}

\subsection{Performance Evaluation}

\subsubsection{Temporal and Spatial Consistency Analysis}
Figure~\ref{fig_data_Aug} illustrates the \gls{cdf}
of the \gls{mape} between consecutive frames in the training dataset, evaluated at different sampling rates with and without image augmentation.
As shown in Fig.~\ref{fig_Wo_Aug}, when no image augmentation is applied, decreasing the sampling rate $\tau_o$ leads to a lower temporal correlation between consecutive frames. 
This indicates that lower sampling rates promote greater temporal variation between frames, which is desirable for training the proposed \gls{ts-jepa}, as it encourages the learning of meaningful temporal dynamics rather than repetitive transitions.
If the sampling rate is too high, consecutive frames become nearly identical, making it easier for the predictor to learn identity mappings, limiting the encoder’s ability to extract semantically rich temporal representations.

Moreover, the application of image augmentation significantly increases the range of \gls{mape} values, as seen in Fig.~\ref{fig_W_Aug}, with up to a five-fold increase compared to the one without image augmentation.
This reflects improved spatial diversity across frames, which is critical for learning robust and generalizable embeddings. 
Therefore, the combination of a well-chosen sampling rate and image augmentation ensures that the frames fed into \gls{ts-jepa} are both temporally and spatially diverse.
This diversity is key to effectively capture semantic representations and latent system behaviors within the embedding space, thereby improving the robustness and generalization capability of the proposed \gls{ts-jepa} across varying control environments.

\begin{figure}
    \centering
    \subfigure[Auto-encoder model.\label{fig_t_SNE_baseline}]{\includegraphics[width=0.24\textwidth]{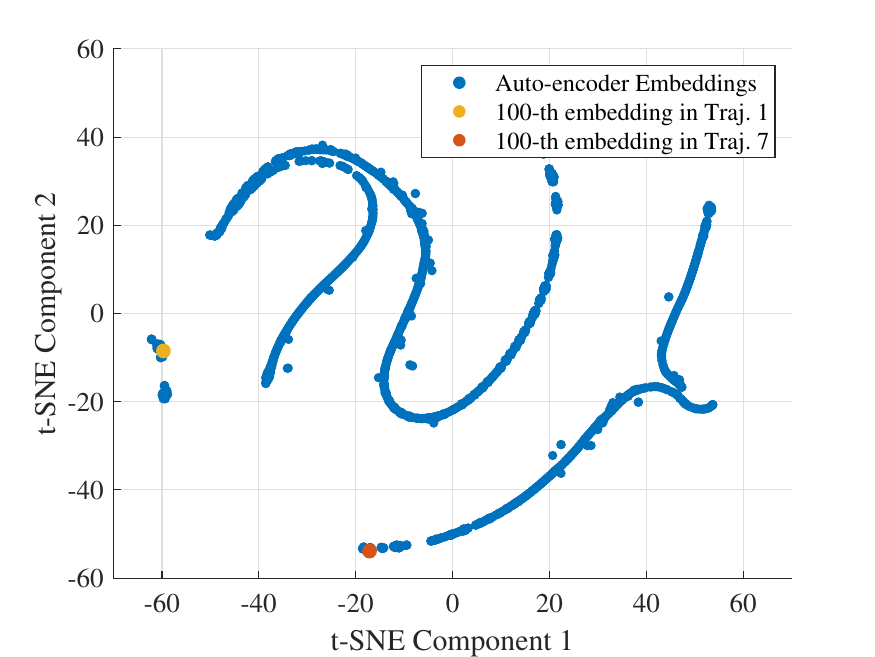}}
    \subfigure[TS-JEPA model.\label{fig_t_SNE_proposed}]{\includegraphics[width=0.24\textwidth]{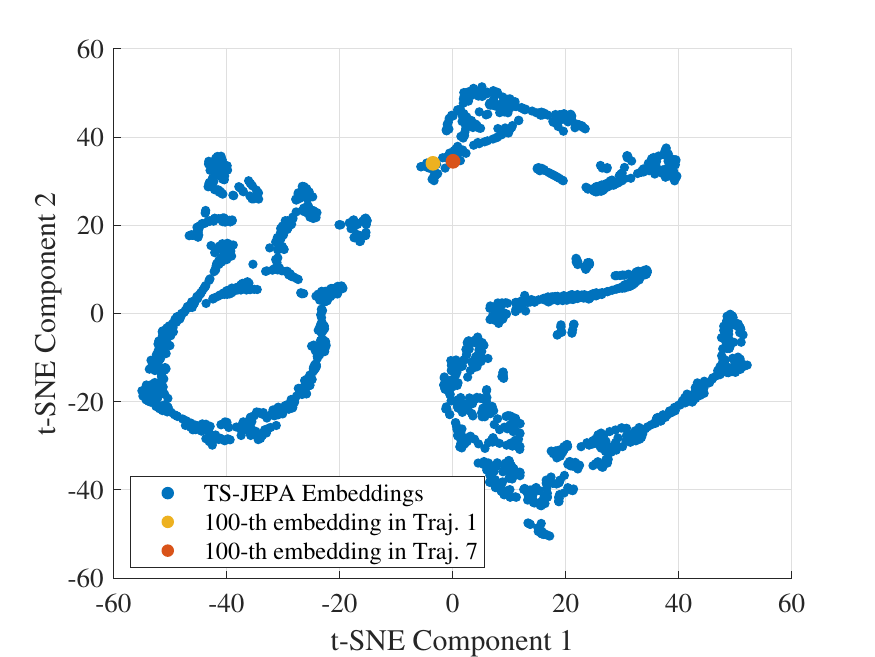}} 
    \subfigure[Frames in different trajectories.\label{fig_frames}]{\includegraphics[width=0.49\textwidth]{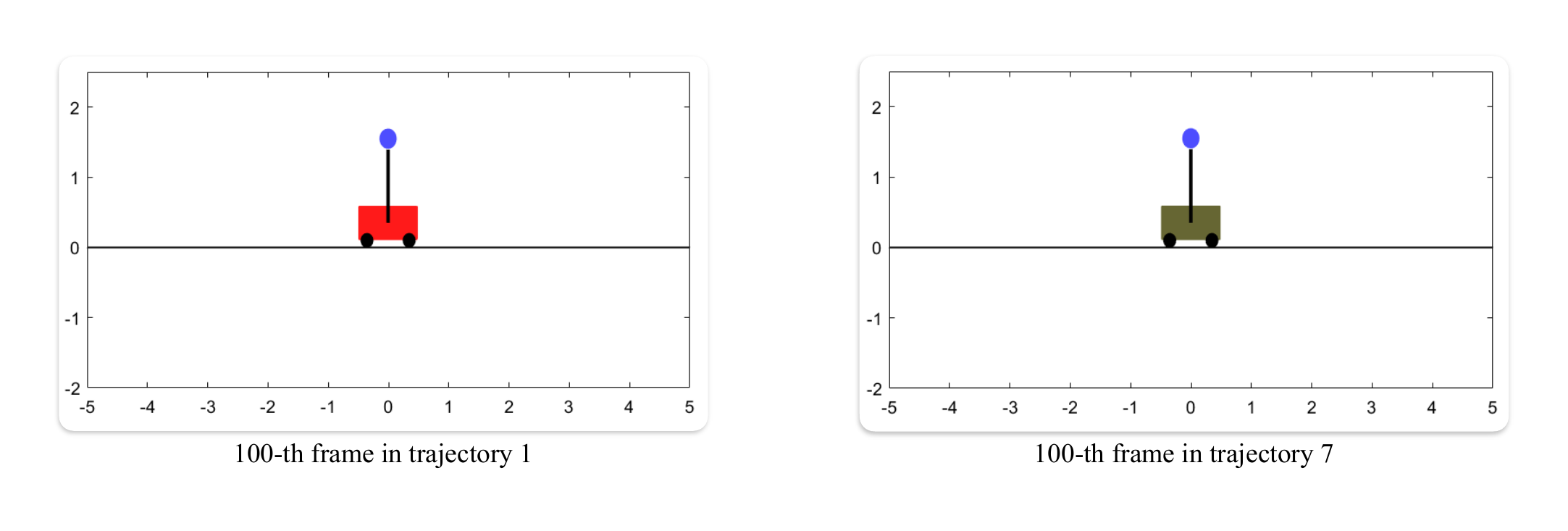}} 
   \caption{t-Distributed stochastic neighbor embedding (t-SNE) visualization of learned embeddings from twenty trajectories: (a) generative auto-encoder model and (b) proposed TS-JEPA model. Sub-figure (c) presents an example of two distinct frames from different trajectories whose embeddings appear in (a) and (b).}
    \label{fig_t_SNE}
\end{figure}

\subsubsection{Encoder Performance}
Fig.~\ref{fig_t_SNE} presents the \gls{t-sne} visualization of learned embeddings from twenty trajectories, comparing the context encoder of the proposed \gls{ts-jepa} with the encoder of a generative auto-encoder model. 
The embeddings produced by the proposed \gls{ts-jepa} exhibit distinct and well-clustered structures, indicating that it effectively captures meaningful semantic embeddings of the device states.
In contrast, the embeddings generated by the generative auto-encoder model exhibit weaker clustering and greater overlap, indicating a limited ability to distinguish between semantically different device states within embedding space.

To further evaluate the semantic encoding capability of the proposed \gls{ts-jepa}, we highlight embeddings corresponding to two different frames (with varying cart colors) that share identical device states (cart location and pendulum angle), as shown in Fig.~\ref{fig_frames}. 
The proposed \gls{ts-jepa} maps these frames to nearby points in the embedding space, indicating that it captures the underlying semantic embeddings within the frames irrespective of visual differences. 
Conversely, the generative auto-encoder model maps them to distant embeddings, revealing its sensitivity to superficial variations.
These results demonstrate the robustness of the proposed \gls{ts-jepa} in learning semantically invariant embeddings, which is crucial for control tasks. 
By directly leveraging these low-dimensional embeddings for control command prediction, the proposed \gls{ts-jepa} avoids the need to reconstruct high-dimensional states, as required by generative models, achieving computational efficiency and improved performance in downstream control applications.

\begin{figure*}
    \centering
    \subfigure[Command prediction accuracy.\label{fig_control_pred}]{\includegraphics[width=0.25\textwidth]{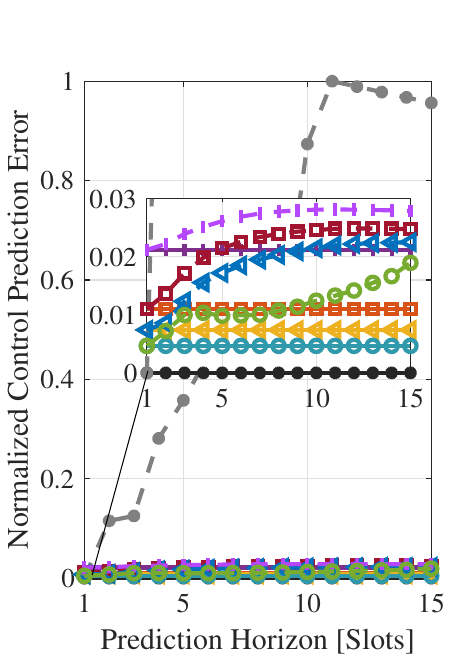}}
    \subfigure[Communication cost.\label{fig_comm_bits}]{\includegraphics[width=0.25\textwidth]{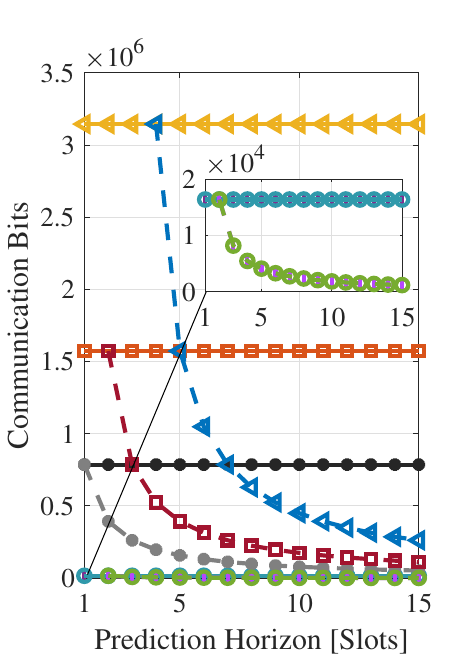}} 
    \subfigure[Control accuracy.\label{fig_norm_score}]{\includegraphics[width=0.48\textwidth]{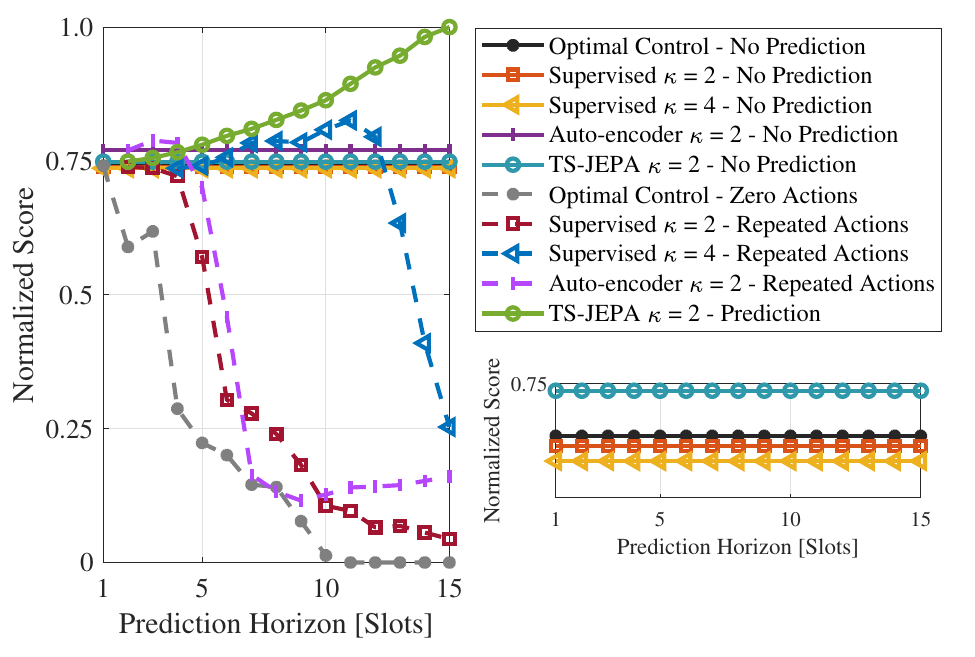}} 
   \caption{Comparison between the proposed semantic-driven predictive control and different control baselines in a special case of a single device in terms of (a) command prediction accuracy, (b) communication cost, and (c) control accuracy.}
    \label{fig_ts-jepa_pred}
\end{figure*}

\subsubsection{Encoding and Prediction Capability Evaluation}
Fig.~\ref{fig_ts-jepa_pred} compares the proposed semantic-driven predictive control with several control baseline models in a single-device scenario, evaluating command prediction accuracy, communication cost, and normalized control score across the prediction horizon.
To validate the encoding capability of the proposed approach, we first consider a no-prediction case where the remote controller receives either high-dimensional states or low-dimensional embeddings at every time slot. 
The proposed \gls{ts-jepa} with two consecutive frames ($\kappa=2$) is compared to: (i) a supervised learning model with $\kappa=2$ and $\kappa=4$, (ii) a generative auto-encoder with $\kappa=2$, and (iii) the optimal non-linear control policy. 
As shown in Fig.~\ref{fig_control_pred}, the proposed \gls{ts-jepa} achieves the lowest normalized control prediction error, closely matching the optimal policy while requiring significantly fewer communication bits (Fig.~\ref{fig_comm_bits}). 
This leads to a normalized control score (Fig.~\ref{fig_norm_score}) that is nearly optimal, validating the \gls{ts-jepa} encoder's ability to extract semantic embeddings from high-dimensional frames and support efficient remote control via the semantic actor model.

In contrast, the generative auto-encoder model exhibits higher control prediction error due to reconstruction inaccuracies and limited temporal data, although its score surpasses that of the supervised learning baseline. 
This is attributed to its use of a non-linear control policy during control command computation.
Meanwhile, the supervised model suffers from poor generalization due to limited training diversity, requiring more high-dimensional frames for improved prediction.
Increasing $\kappa$ from $2$ to $4$ enhances control performance but incurs higher communication costs, emphasizing the trade-off between communication overhead and control performance in generative approaches. 

To evaluate the prediction capability of the proposed \gls{ts-jepa}, we evaluate a case where the device transmits only at the initial time slot, and the remote controller needs to predict future control commands.
In this case, the proposed \gls{ts-jepa} is compared to the control baseline models with repetitive-action and zero-action strategies. 
The results show that the proposed \gls{ts-jepa} maintains the lowest control prediction error throughout the prediction horizon, outperforming other control baseline models whose errors grow due to the lack of dynamic inference.
This is enabled by \gls{ts-jepa}’s ability to predict future embeddings using its predictor and compute control commands using the semantic actor model.
Although \gls{ts-jepa}’s prediction error increases slightly over time due to auto-regressive accumulation, it achieves better normalized scores than all baselines, including the optimal policy.
This is due to its aggressive control behavior, which rapidly drives the device toward the desired state. 
Moreover, the proposed \gls{ts-jepa} achieves these results with significantly fewer communication bits, demonstrating its effectiveness in reducing communication overhead without compromising control performance.
These results underscore the strength of the proposed \gls{ts-jepa} in encoding and predicting semantic embeddings, enabling scalable and communication-efficient control, particularly in bandwidth-constrained wireless control systems.

\begin{figure*}
    \centering
    \subfigure[Command prediction accuracy.\label{fig_control_pred_Enc_op}]{\includegraphics[width=0.25\textwidth]{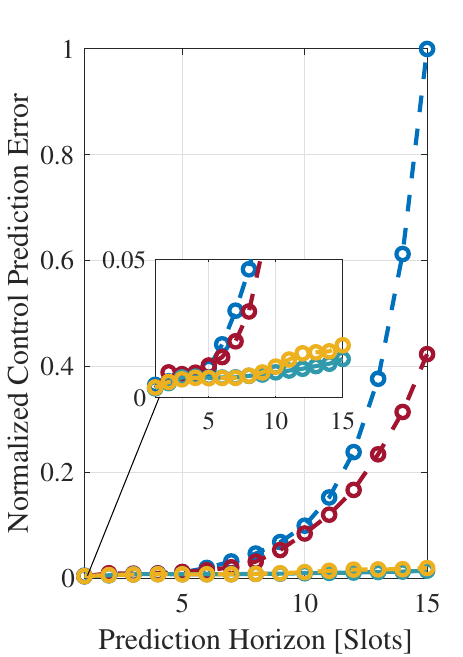}}
    \subfigure[Communication cost.\label{fig_comm_bits_Enc_op}]{\includegraphics[width=0.25\textwidth]{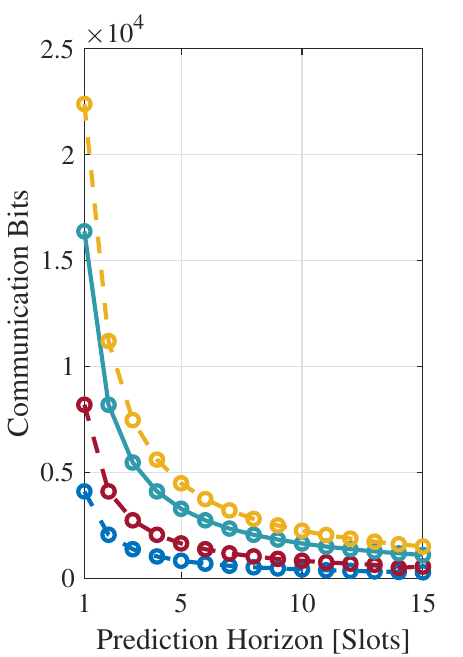}} 
    \subfigure[Control cost.\label{fig_norm_score_Enc_op}]{\includegraphics[width=0.418\textwidth]{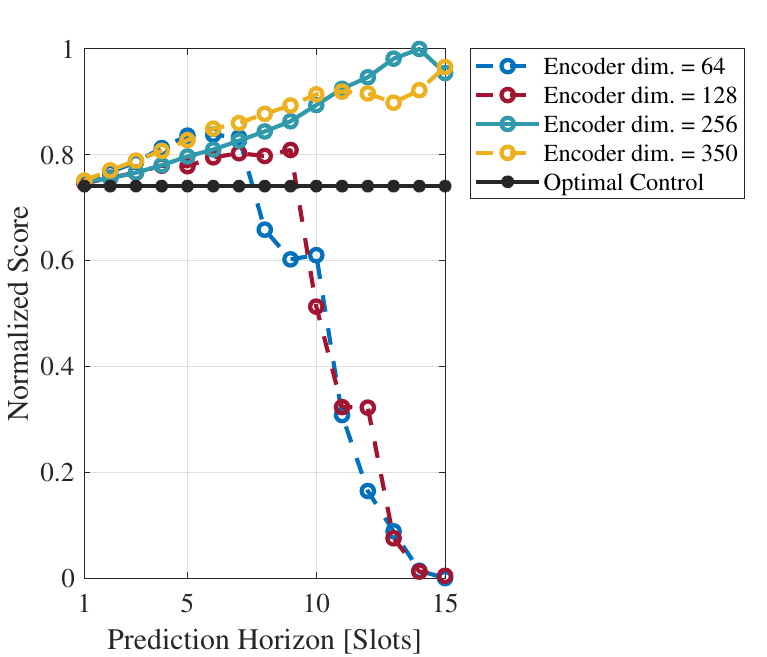}} 
   \caption{Embedding dimensions impact for a single device in terms of (a) command prediction accuracy, (b) communication cost, and (c) control cost.}
    \label{fig_ts-jepa_pred_Enc_op}
\end{figure*} 

\subsubsection{Embedding Dimension Evaluation}
Fig.~\ref{fig_ts-jepa_pred_Enc_op} investigates the impact of the embedding dimension on the performance of the proposed semantic-driven predictive control in a single-device scenario, evaluating command prediction accuracy, communication cost, and control performance across the prediction horizon.
As shown in Fig.~\ref{fig_control_pred_Enc_op}, increasing the dimensionality of the embedding significantly reduces the normalized control prediction error, highlighting the importance of higher-dimensional embeddings in capturing richer semantic features from high-dimensional frames, which improves the accuracy of downstream control command prediction.
However, as shown in Fig.~\ref{fig_comm_bits_Enc_op}, this improvement in semantic representations comes at the cost of increased communication overhead. 
Larger semantic embeddings require more communication bits to transmit, which is a critical consideration under limited network capacity.
Nevertheless, Fig.~\ref{fig_norm_score_Enc_op} demonstrates that higher-dimensional embeddings also prolong the normalized control score over the prediction horizon, indicating a more stable control performance.
These results highlight a fundamental trade-off between larger embedding dimensions to enhance prediction accuracy and control robustness, which also increases communication cost.
Therefore, careful selection of the embedding dimension is essential to balance between communication efficiency and control accuracy, ensuring the semantic-driven predictive control remains scalable and resource-efficient.

\subsubsection{Training Dataset size Evaluation}
\begin{figure*}
    \centering
    \subfigure[Command prediction accuracy.\label{fig_control_training_size}]{\includegraphics[width=0.25\textwidth]{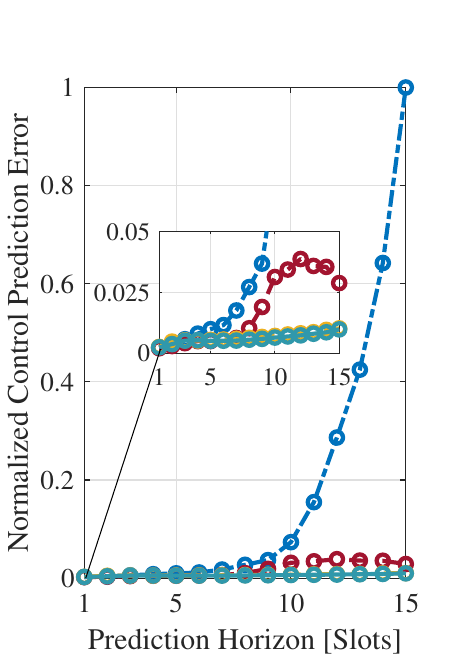}}
    \subfigure[Communication cost.\label{fig_comm_bits_training_size}]{\includegraphics[width=0.25\textwidth]{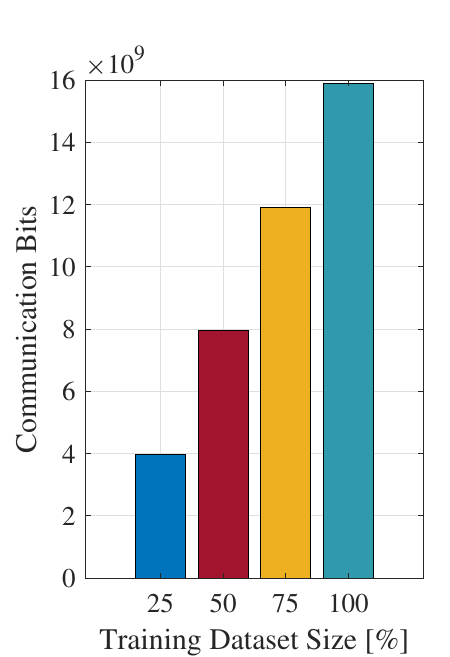}} 
    \subfigure[Control cost.\label{fig_norm_score_training_size}]{\includegraphics[width=0.418\textwidth]{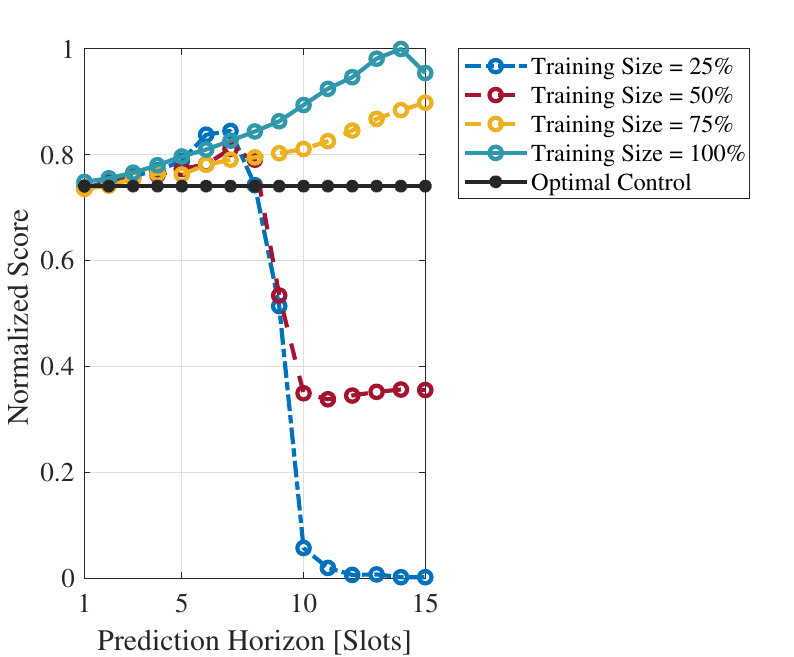}} 
   \caption{Training Dataset size impact in a single device in terms of (a) command prediction accuracy, (b) communication cost, and (c) control accuracy.}
    \label{fig_ts-jepa_pred_training_size}
\end{figure*}

Fig.~\ref{fig_ts-jepa_pred_training_size} evaluates the impact of training dataset size on the performance of the proposed semantic-driven predictive control in a single-device scenario, focusing on command prediction accuracy, communication cost, and control performance over the prediction horizon. 
As illustrated in Fig.~\ref{fig_control_training_size}, increasing the training dataset size significantly improves the normalized control prediction accuracy.
This improvement indicates that a sufficiently large dataset is essential to capture the underlying latent dynamics critical to downstream control command prediction.
However, as shown in Fig.~\ref{fig_comm_bits_training_size}, this enhanced accuracy comes at the cost of a higher communication overhead. 
Larger training datasets typically lead to richer semantic embeddings, which may require more bits to transmit, posing a challenge in resource-constrained wireless control systems.
In contrast, training with insufficient data tends to underfit, leading to noisy semantic embeddings and poor generalization. 
In the context of embedding prediction, limited training data can also cause representation collapse, degrading semantic richness and control robustness.
Despite the communication cost, Fig.~\ref{fig_norm_score_training_size} demonstrates that training with larger datasets maintains a higher normalized control score over extended prediction horizons, reflecting improved long-term control performance and predictive robustness.
In contrast, smaller training datasets result in a rapid decline in the normalized control score, signaling deterioration in control performance under long-horizon prediction due to poor generalization.
These results highlight a key trade-off: while increasing training dataset size improves control accuracy and robustness, it also increases communication overhead. 
Therefore, careful selection of the training dataset size is necessary to balance between control performance and communication efficiency. 
Optimizing this trade-off is essential for ensuring the scalability and practicality of the proposed semantic-driven predictive control in large-scale, bandwidth-constrained wireless networked control systems.

\begin{figure}
    \centering
    \subfigure[Command prediction accuracy.\label{fig_action_snr}]{\includegraphics[width=0.24\textwidth]{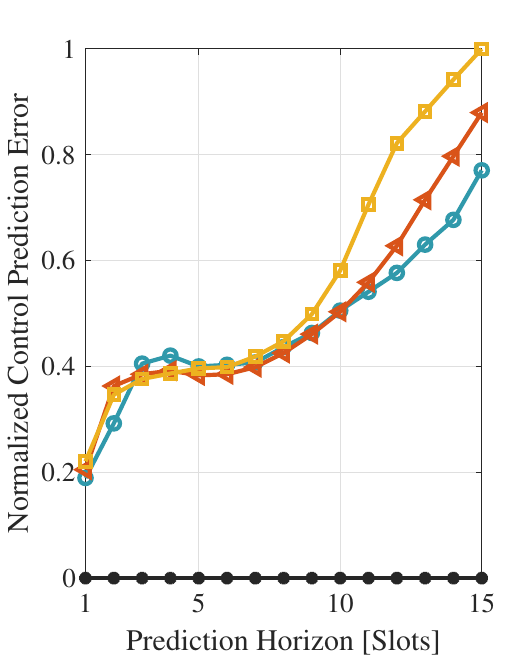}}
    \subfigure[Control cost.\label{fig_score_snr}]{\includegraphics[width=0.24\textwidth]{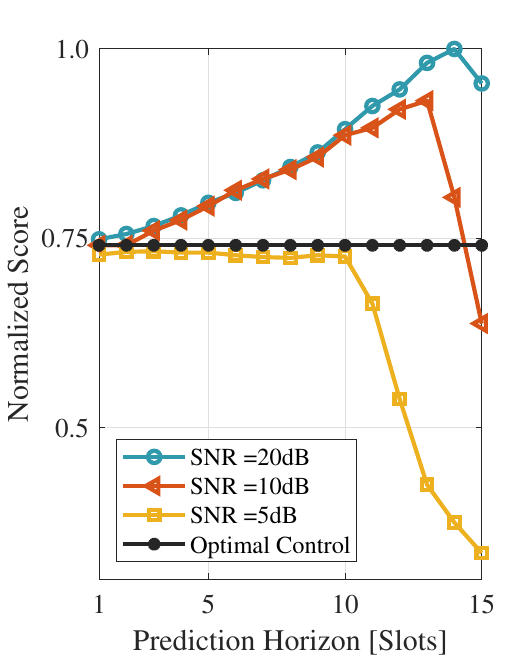}} 
   \caption{The impact of target signal-to-noise ratio value in a single device in terms of : (a) command prediction accuracy and (b) control cost.}
    \label{fig_SNR_impact}
\end{figure} 

\subsubsection{Target SNR Evaluation}
Fig.~\ref{fig_SNR_impact} evaluates the impact of target \gls{snr} values on the performance of the proposed semantic-driven predictive control in a single-device scenario, focusing on command prediction accuracy and control performance over the prediction horizon.
As illustrated in Fig.~\ref{fig_action_snr}, 
lowering the target \gls{snr} from $20 \,\mathrm{dB}$ to $5 \,\mathrm{dB}$ significantly degrades the control prediction accuracy. 
This degradation arises because lower \gls{snr} values result in fewer successfully received samples during training, limiting the \gls{ts-jepa} model’s ability to capture the underlying latent dynamics necessary for accurate control command prediction.
Correspondingly, Fig.~\ref{fig_score_snr} shows a decline in normalized control scores under low \gls{snr} values, further confirming the negative impact on long-horizon control performance.
Fewer training samples impair the model’s generalization capability, leading to reduced predictive accuracy and higher normalized control prediction error over time.
In contrast, training the \gls{ts-jepa} under high \gls{snr} values captures the underlying latent dynamics, enabling more robust and accurate control over extended horizons. 
This results in a higher normalized control score, reflecting robust prediction accuracy and effective control over extended prediction horizons.
However, achieving high \gls{snr} value requires increased wireless resource consumption, highlighting a key trade-off: improving predictive control robustness versus limiting communication cost. 
These findings emphasize the importance of considering \gls{snr} values during model training, particularly in bandwidth-constrained wireless control systems.

\begin{figure}
    \centering
    \subfigure[Proposed TS-JEPA model.\label{fig_ts-jepa_scheduling}]{\includegraphics[width=0.24\textwidth]{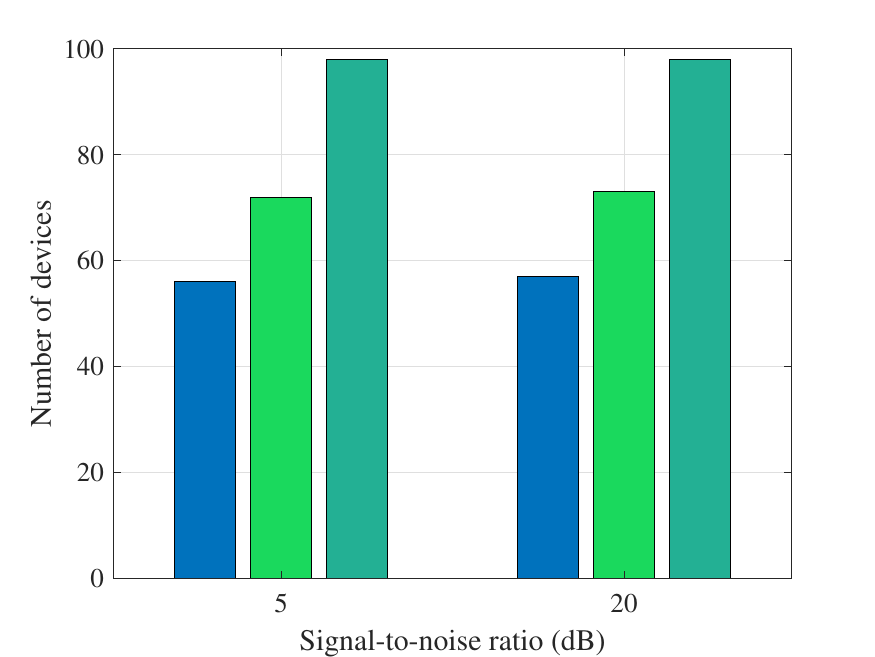}}
    \subfigure[Supervised learning model.\label{fig_baseline_scheduling}]{\includegraphics[width=0.24\textwidth]{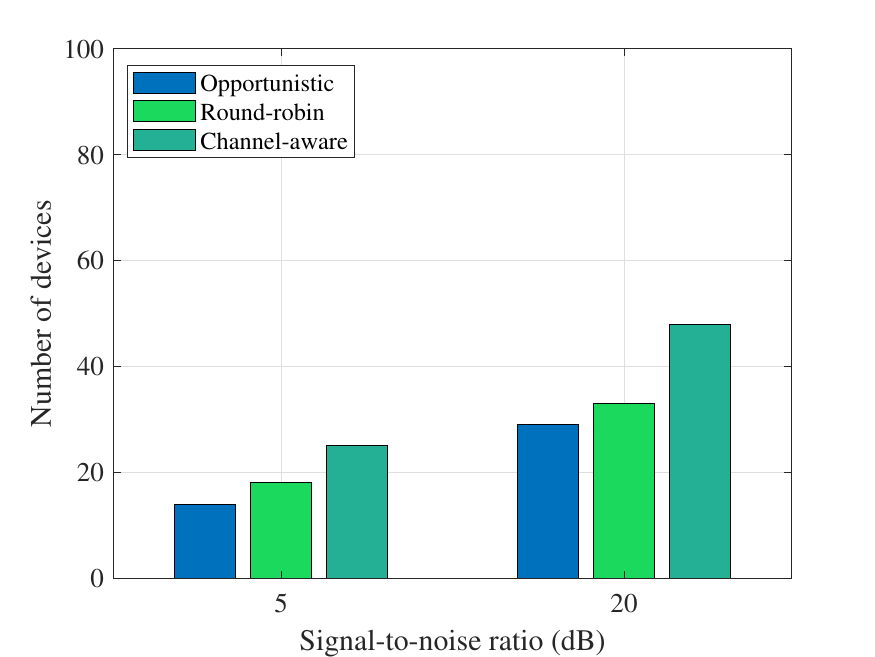}} 
   \caption{Total number of devices under different scheduling approaches and target signal-to-noise ratio values for the: (a) proposed TS-JEPA model and (b) supervised learning model.}
    \label{fig_scheduling}
\end{figure} 

\subsubsection{System scalability with encoding capability}
Fig.~\ref{fig_scheduling} presents the scalability performance of the proposed channel-aware scheduling compared to baseline scheduling approaches.
The evaluation examines the maximum number of devices that can be supported while maintaining acceptable control performance, defined as a normalized score within the specified range $ \left[ 0.74, 1.0 \right]$, across different \gls{snr} values for the proposed \gls{ts-jepa} and supervised learning models. 
The results demonstrate that the proposed scheduling, when combined with the \gls{ts-jepa}, significantly outperforms round-robin and opportunistic baselines in terms of scalability under different \gls{snr} values.
This superior performance is attributed to two key design features. 
First, the \gls{ts-jepa} effectively compresses high-dimensional states into low-dimensional semantic embeddings, reducing transmission overhead and enabling support for a larger number of devices. 
Second, channel-aware scheduling dynamically selects devices for uplink transmission based on their \gls{aoi} and channel conditions, ensuring timely and reliable updates.
In contrast, round-robin scheduling offers moderate scalability due to its fairness in transmission updates, but suffers from inefficiencies caused by ignoring channel conditions and device urgency. 
The opportunistic scheduling, despite leveraging channel conditions, exhibits the lowest scalability.
This is attributed to its tendency to favor devices with good channels, regardless of their update criticality, resulting in poor control performance and sub-optimal resource utilization in large-scale deployment.
Moreover, the comparison between the \gls{ts-jepa} and supervised learning models highlights the robustness of the proposed semantic-driven control under varying \gls{snr} values.   
The \gls{ts-jepa} model's semantic embeddings allow for efficient operation even under low \gls{snr} values, while the supervised baseline model struggles due to the need to transmit high-dimensional states, consuming more bandwidth and restricting the number of supported devices. 
These results underscore the effectiveness of combining semantic compression with channel-aware scheduling to provide scalable and reliable control in bandwidth-constrained wireless networks.

\begin{figure}
    \centering
    \subfigure[Proposed TS-JEPA model.\label{fig_ts-jepa_scheduling_pred}]{\includegraphics[width=0.24\textwidth]{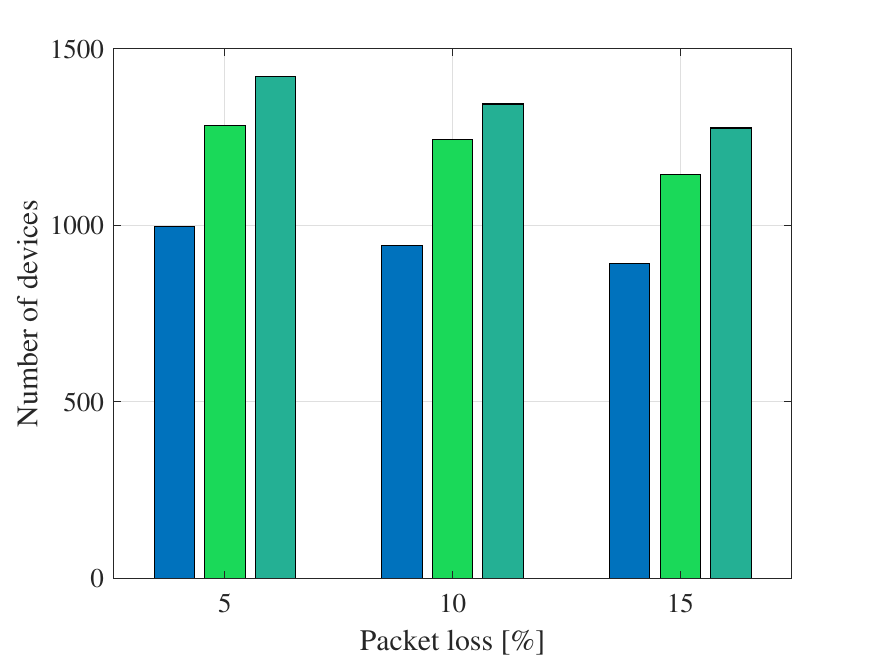}}
    \subfigure[Supervised learning model.\label{fig_baseline_scheduling_pred}]{\includegraphics[width=0.24\textwidth]{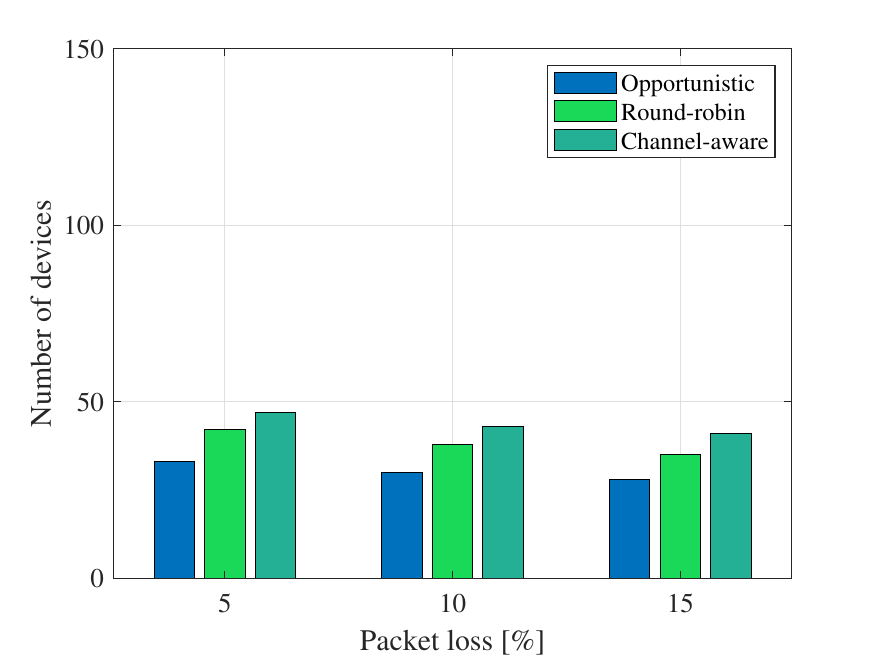}} 
   \caption{Total number of devices under different scheduling approaches and packet losses for the: (a) proposed TS-JEPA model and (b) supervised learning model.}
    \label{fig_scheduling_pred}
\end{figure}

\subsubsection{System scalability with prediction capability}
Fig.~\ref{fig_scheduling_pred} evaluates the scalability of the proposed channel-aware scheduling against baseline approaches under varying packet losses.
The evaluation focuses on determining the maximum number of devices that can be reliably supported while maintaining acceptable control performance, defined by a normalized score within the specified range $[0.74, 1.0]$, for both the proposed \gls{ts-jepa} and supervised learning models.
The results demonstrate that the proposed framework, combining semantic-driven predictive control with channel-aware scheduling, substantially outperforms round-robin and opportunistic scheduling approaches with the supervised learning model in terms of scalability, even under adverse network conditions.
This superior performance is attributed to two design features. 
First, \gls{ts-jepa} encodes high-dimensional states into low-dimensional semantic embeddings, significantly reducing communication overhead and enabling the support of more devices. 
Moreover, its predictive capability allows it to predict future embeddings even when packets are lost, thereby maintaining an acceptable control performance without requiring retransmission. 
Second, channel-aware scheduling dynamically prioritizes transmissions based on device-specific \gls{aoi} and channel conditions, ensuring timely and reliable updates.
In contrast, the supervised learning model exhibits limited robustness and scalability, primarily due to its dependence on transmitting high-dimensional states, which burdens network capacity and lacks predictive capability under packet loss. 
These results highlight the effectiveness of integrating semantic-driven predictive control with channel-aware scheduling to enable robust and scalable control across large-scale wireless networked systems operating under limited network capacity.

\section{Conclusion}
\label{Sec_conclusion}

This work introduced a novel semantic-driven predictive control, integrated with channel-aware scheduling, to address the core challenges of control robustness, communication efficiency, and scalability in wireless networked control systems.
The proposed approach employs a self-supervised \gls{ts-jepa} with a semantic actor model to encode high-dimensional sensory data into low-dimensional semantic embeddings, significantly reducing uplink communication overhead. 
Beyond efficient encoding, \gls{ts-jepa} enables forward prediction of future embeddings, allowing the remote controller to infer future embeddings without continuous uplink updates.
To further enhance communication efficiency, a channel-aware scheduling dynamically prioritizes embedding transmission devices based on channel quality and \gls{aoi}, ensuring timely updates where they are most critical. 
Simulation results on large-scale inverted cart-pole systems demonstrate that the proposed framework not only achieves high control accuracy and low prediction error but also dramatically reduces communication costs compared to conventional baselines. 
Furthermore, the proposed framework supports robust control performance over extended prediction horizons and scales efficiently to accommodate significantly more devices in limited uplink capacity.

\bibliographystyle{IEEEtran}
\bibliography{IEEEabrv,bibliography}
\end{document}